\newcommand{\BP}{$G_{\rm BP}$}
\newcommand{\RP}{$G_{\rm RP}$}
\newcommand{\Gaia}{\emph{Gaia}}
\newcommand{\GDR}{\Gaia~DR}

\newcommand{\footlabel}[2]{%
    \addtocounter{footnote}{1}%
    \footnotetext[\thefootnote]{%
        \addtocounter{footnote}{-1}%
        \refstepcounter{footnote}\label{#1}%
        #2%
    }%
    $^{\ref{#1}}$%
}

\newcommand{\footref}[1]{%
    $^{\ref{#1}}$%
}

\def\teff{$T_\mathrm{eff}$}
\documentclass{aa}  
\usepackage{hyperref}
\usepackage{graphicx}
\usepackage{txfonts}
\usepackage[normalem]{ulem}
\usepackage{cancel}

\begin{document} 

   \title{Gaia Data Release 1}
   
   \subtitle{Principles of the Photometric Calibration of the $G$ band}


   \author{
  J.~M.~Carrasco\inst{1} 
\and  
  D.~W.~Evans\inst{2}
\and
  P.~Montegriffo\inst{3}
\and
  C.~Jordi\inst{1}
\and
  F.~van~Leeuwen\inst{2}
\and
  M.~Riello\inst{2}
\and
  H.~Voss\inst{1}
\and
  F.~De~Angeli\inst{2}
\and
  G.~Busso\inst{2}
\and
  C.~Fabricius\inst{1}
\and
  C.~Cacciari\inst{3}
\and
  M.~Weiler\inst{1}
\and
  E.~Pancino\inst{5,8}
\and
  A.~G.~A.~Brown\inst{6}
\and
  G.~Holland\inst{2}
\and
  P.~Burgess\inst{2}
\and
  P.~Osborne\inst{2}
\and
  G.~Altavilla\inst{3}
\and
  M.~Gebran\inst{7}
\and
  S.~Ragaini\inst{3}
\and
  S.~Galleti\inst{3}
\and
  G.~Cocozza\inst{3}
\and
  S.~Marinoni\inst{4,8}
\and
  M.~Bellazzini\inst{3}
\and
  A.~Bragaglia\inst{3}
\and 
  L.~Federici\inst{3}
\and
  L.~Balaguer-N\'u\~nez\inst{1}
          }

   \institute{
Institut de Ci\`encies del Cosmos, Universitat de Barcelona (IEEC-UB), Mart\'i Franqu\`es 1, E-08028 Barcelona, Spain\\
\email{carrasco@fqa.ub.edu}
\and
Institute of Astronomy, University of Cambridge, Madingley Road, Cambridge
CB3 0HA, UK
\and
INAF -- Osservatorio Astronomico di Bologna, via Ranzani 1, 40127 Bologna, Italy
\and
INAF -- Osservatorio Astronomico di Roma, Via Frascati 33, I-00040, Monte Porzio Catone (Roma), Italy
\and
INAF -- Osservatorio Astrofisico di Arcetri, Largo Enrico Fermi 5, 50125 Firenze, Italy
\and
Sterrewacht Leiden, Leiden University, PO Box 9513, 2300 RA Leiden, the Netherlands
\and
Department of Physics and Astronomy, Notre Dame University-Louaize, PO Box 72, Zouk Mika\"{e}l, Lebanon
\and
ASI Science Data Center, via del Politecnico SNC, I-00133 Roma, Italy
             }

   \date{Received July 4, 2016; accepted October 30, 2016}

 
  \abstract
   {{\Gaia} is an ESA cornerstone mission launched on 19 December 2013 aiming to obtain the most complete and precise 3D map of our Galaxy by observing more than one billion sources.
This paper is part of a
series of documents explaining the data processing and its results for {\Gaia} Data Release 1, focussing on the $G$ band photometry.
}
   {
This paper describes the calibration model 
of the {\Gaia} photometric passband for {\Gaia} Data Release 1.  
   }
   {
   The overall principle of splitting the process into internal and external calibrations is outlined. In the internal
   calibration, a self-consistent photometric system is generated. Then, the external calibration provides the link to
   the absolute photometric flux scales.
   }
   {The {\Gaia} photometric calibration pipeline explained here was applied to the first data release with good results. Details are given of the 
   various calibration elements including the mathematical formulation of the models used and of the extraction and preparation of the required input parameters (e.g. colour terms).
The external calibration in this first release provides the absolute zero point and photometric transformations from the {\Gaia} $G$ passband to other common photometric
   systems.}
   {This paper describes the photometric calibration implemented for the first {\Gaia} data release and the instrumental effects taken into account. 
   For this first release no aperture losses, radiation damage, and other second-order effects have not yet been implemented in the calibration. 
   }

   \maketitle
%


\section{Introduction}
\label{sec:intro}

The {\Gaia} mission \citep{GaiaMission} launched in 2013 by the European Space Agency (ESA) is a global astrometric mission consisting of two
telescopes sharing the same focal plane, continuously scanning the full sky for a minimum of 5 years from the
Sun-Earth Lagrangian L2 point. The {\Gaia} mission will determine positions, parallaxes and proper motions for $\sim1$\% of the Milky Way 
(more than $10^9$ sources).

In addition to astrometric
information, the first {\Gaia} Data Release (\citealt{GDR1}, hereafter {\GDR1}) also contains {\Gaia} mean fluxes obtained in the astrometric field (see
\citealt{PhotTopLevel} for details) observed during the first 14 months of the nominal mission. 
This paper describes the calibration model used in the photometric calibration pipeline (PhotPipe) for {\GDR1}. 
The practicalities for the application of this model and the processing issues found are explained in
\cite{PhotProcessing}. The results of the photometric calibration and the verification process applied are discussed in \cite{PhotValidation}.

\begin{figure*} 
 \centering
 \includegraphics[scale=0.6]{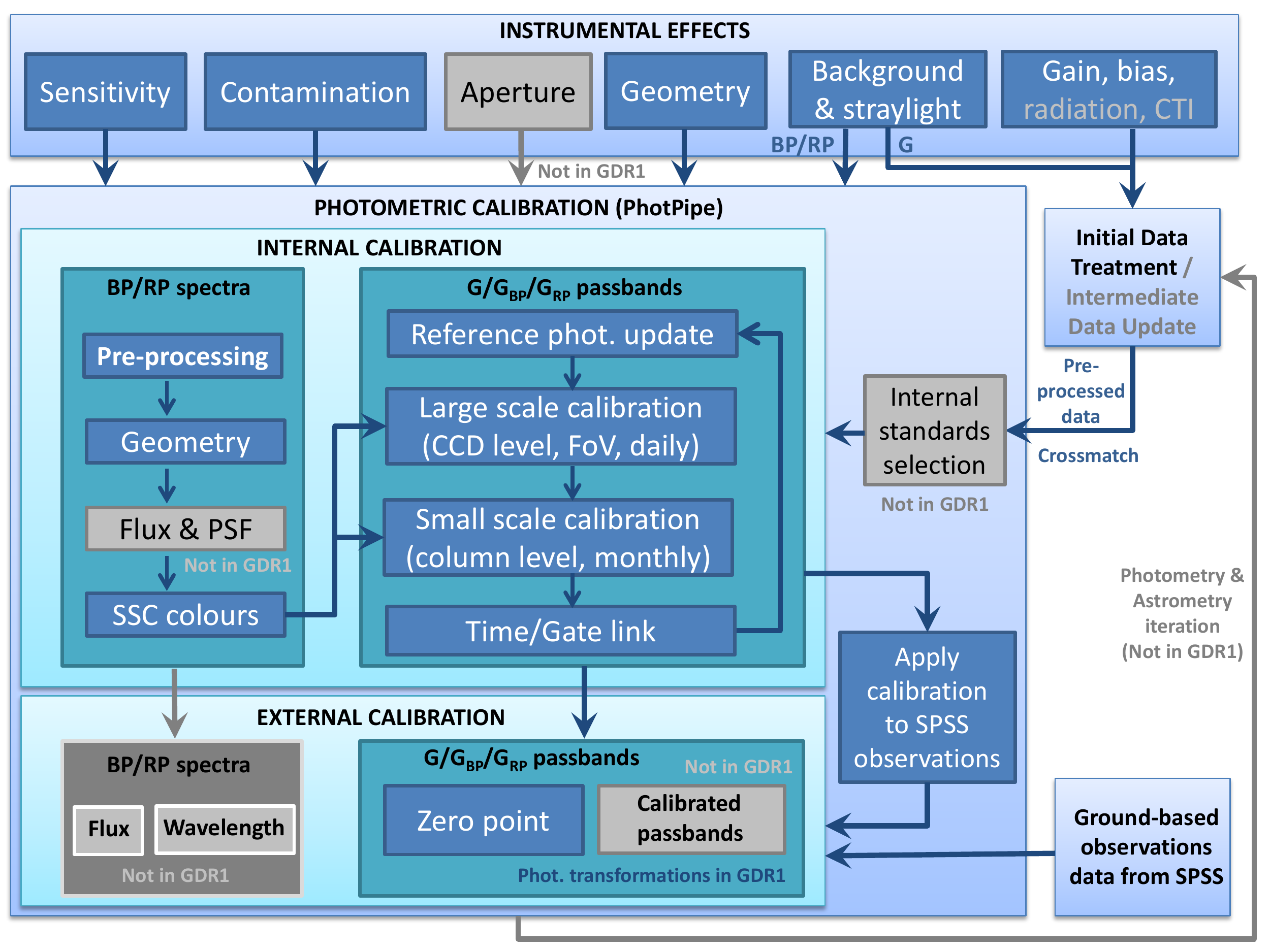}
\caption{
Summary of the instrumental effects and processes present in the {\Gaia} photometric calibration. Grey boxes/arrows indicate those components not considered in
{\GDR1} (to be activated in future releases).
\label{Fig:IterationPlot}
}
\end{figure*}

Figure~\ref{Fig:IterationPlot} provides an overview of the photometric calibration process explained in this paper (see Sect.~\ref{sec:principles}). Grey boxes are used to
indicate processes that were not active in the preparation of {\GDR1}.

The large size of the {\Gaia} focal plane (see Sect.~\ref{sec:instrument} for an overview of the {\Gaia} instrument), the long duration of the space mission (with expected
ageing effects due to radiation), the
observation of a huge number of sources of different type (e.g. different stellar populations, extragalactic sources, 
solar system bodies) 
covering the entire sky with the same instrument, and the need to reach an
accuracy close to the signal-to-noise ratio (S/N) limit make 
the photometric calibration of the {\Gaia} data a challenging task.
For this reason, 
{\Gaia} uses a self-calibration approach
\citep{jordi2006,jordi2007}, similar to \textit{\"ubercalibration} \citep{padmanabhan08},
splitting the problem into internal or relative calibrations (using only {\Gaia} observations) and external calibrations (using a relatively small number of ground-based 
standard stars) 
calibrations.

The goal of the internal calibration (Sect.~\ref{sec:internal}) is to provide an internally consistent flux scale throughout the mission, across the focal plane, and for
all sources (including the full range of brightness and stellar types).  The internal calibration is carried out by grouping observations taken in the same instrumental
configuration (same telescope or field of view, FoV; CCD\footnote{Charge coupled device.}; pixel column\footnote{We use ``column'' for the set of 4500 pixels in the CCD
aligned with the scanning direction and ``line'' for the set of 1966 pixels aligned in the direction across this scanning direction (see Fig.~\ref{fig:focalplane}).};
effective exposure time) and in the same time range, for time-dependent effects.  Each of these configurations represents a calibration unit (CU) (see
Sect.~\ref{sec:instrument}.  The passband variations across the focal plane are modelled with a colour dependency (see Sect.~\ref{sec:internal}). The extraction and
application of the required colour information into the internal calibration model is described in Sect.~\ref{sec:colour}.

The internal calibration model assumes that the true set of 
calibration parameters can be derived by comparing
the observations of a large set of internal reference sources (isolated and non-variable) with their predicted observations.
The prediction is made
from the {\Gaia} internal reference fluxes themselves (see Sect.~\ref{sec:update} where the definition of the mean
photometry per source is provided) and the current calibration parameters. The instrumental variations can be described as
small corrections to the current set of parameters. 

Once the internal calibration is applied, the external calibration can be performed (see Sect.~\ref{sec:external}). The aim of the external calibration
is to determine the characteristics of the mean instrument (true 
passbands) by using a suitable number of
spectro photometric standard stars (SPSS) whose absolute spectral energy distribution (SED) is known with great
accuracy from ground-based observations \citep{pancino2012,altavilla2015,marinoni2016}.
In {\GDR1}, the internally calibrated fluxes (transformed to $G$ magnitudes) and zero point plus photometric transformations to other common photometric systems are provided. 

Finally, in Sect.~\ref{sec:conclusions} a brief summary is presented.

\section{Defining the reference system}
\label{sec:principles}

As mentioned in the Introduction,
{\Gaia} photometry is self-calibrated. 
The overall principle of the calibrations is that of bootstrapping. Only data from the satellite is used in the internal calibration to
define the reference fluxes.

Common approaches of the photometric calibration used for small focal planes are not suitable for {\Gaia}. The use of a few very reliable standard sources would require
perfect knowledge of the bandpasses and the stellar SEDs to avoid introducing systematic effects.  Most of the best available catalogues only cover a small part of the sky
and contain an insufficient number of sources. Even the best catalogues have a lower photometric accuracy and angular resolution than needed to calibrate the {\Gaia} data.
Furthermore, ground-based standard source catalogues usually only contain bright stars and cannot be applied to the magnitude range covered by {\Gaia}.  Finally, {\Gaia} is
not a pointing observatory and no specific calibration operations are programmed in order to repeatedly observe a list of standards (as can be done for instance for the
Hubble Space Telescope, HST, \citealt{bohlin16}).

In 2006, when the {\Gaia} Data Processing and Analysis Consortium (DPAC) was created to deal with the instrument calibration, we had already foreseen a different approach
(see \citealt{jordi2006,jordi2007}). The process we considered at that time splitting internal and external calibrations has been used in other projects and has been called
\textit{\"ubercalibration} \citep{padmanabhan08}. These ideas are not new to optical astronomy; precursors may be found in the work of \cite{maddox1990},
\cite{honeycutt1992}, \cite{fong1992}, \cite{manfroid1993}, \cite{fong1994}, and \cite{glazebrook1994}.  All surveys using large focal planes, for example SDSS in its 8th
release \citep{aihara2011}, the Euclid mission \citep{markovic16}, MegaCam \citep{MegaCam}, PanSTARRS \citep{panstarrs,panstarrsSDSS}, and DES \citep{DES}, use similar
approaches.

This \textit{\"ubercalibration} procedure performs a self-consistent likelihood maximisation of the 
magnitudes of non-variable sources and assumes a high degree of overlapping because these calibration sources have been observed under many different
instrumental conditions, ranges of time, and all across the sky. This procedure greatly increases the number and types of 
reference stars available without needing perfect knowledge of their SED.

Unlike the model explained in Sect.~\ref{sec:internal}, no colour term is used in \cite{padmanabhan08} and the authors warn in their conclusions that some colour
dependencies are present in their results\footnote{SDSS filters are narrower than in the {\Gaia} case, so the effect of the colour terms is expected to be smaller than 1\%
for a wide range of objects. SDSS calibration relies on mid-colour range stars from the main sequence, so that filter zero points are well-defined. As a consequence, colour
terms may be applied with good accuracy.}. In order to avoid these colour biases we introduce the dependency with the colour of the source.

During the internal calibration, the reference fluxes\footnote{Contrary to \cite{padmanabhan08} who use magnitudes, we carry out the photometric calibration in fluxes to
avoid biases caused by the non-linear transformation between the two, which amounts to 8 mmag at G=20.} for the internal standard sources need to be set up. This is done
using an iterative scheme\footnote{\cite{padmanabhan08} perform a global fit of the calibration parameters, instead of an iterative scheme such as ours.}, illustrated in
Fig.~\ref{Fig:IterationPlot} in the box for the internal calibration of $G$, {\BP} and {\RP} (see Sect.~\ref{sec:instrument} for the description of the components of the
instrument).

As {\Gaia} is a self-calibrated instrument using millions of internal reference sources and because their fluxes must also be determined during the process, the internal
instrument calibration (Sect.~\ref{sec:internal}) and the reference photometry update (Sect.~\ref{sec:update}) are closely linked. With so many (millions) reference sources
observed several times and so many CUs (Table~\ref{tab:NCU}) to be calibrated, a global fit would require a very large matrix to be solved as a whole. We preferred reducing
the dimensionality of the problem by splitting the process into independent CU calibrations. Moreover, the necessary introduction of the colour of the sources in the
equations imposes an iterative approach.

\begin{table}
\begin{center}
\caption{Number of CUs, $N_{\rm CU}$, used for the {\GDR1} photometric calibration. $N_{\rm CU}$ is the product of all the other columns (number of CCD rows and strips,
gate/WC configuration, FoV, AC bin, and time
range). See Sect.~\ref{sec:instrument} for the description of the instrument and for the definition of a CU.}
\tiny
\begin{tabular}{p{0.7cm}p{0.5cm}p{0.5cm}p{0.5cm}p{0.7cm}p{0.5cm}p{0.5cm}p{0.5cm}|p{1.0cm}}
\hline
Instrum. & Scale & $N_{\rm rows}$ & $N_{\rm strips}$ & $N_{\rm gate/WC}$ &  $N_{\rm FoV}$ & $N_{\rm AC}$ & $N_{\rm time}$ & $N_{\rm CU}$\\
\hline
AF       & LS & 7 & 8/9 & 10 & 2 &  -  & 420 & 529\ 200 \\
AF       & SS & 7 & 8/9 & 10 & - & 492 &  1  & 309\ 960 \\
BP/RP    & LS & 7 & 1 &  6 & 2 &  -  & 420 &  35\ 280 \\
BP/RP    & SS & 7 & 1 &  6 & - & 492 &  1  &  20\ 664 \\
\hline          
\end{tabular}
\label{tab:NCU}
\end{center}
\end{table}

In the initial run of the process shown in Fig.~\ref{Fig:IterationPlot}, all the weighted means from all the uncalibrated observations\footnote{Fluxes are derived with a
PSF/LSF (point and line spread function, respectively; see Sect.~\ref{sec:instrument}) fitting process.} for all sources are computed to act as a first set of reference
fluxes. These approximate reference fluxes can then be used to produce a rough set of calibration parameters. Using these first parameters, a better set of reference fluxes
can be generated. Better calibration parameters and reference fluxes are produced during each iteration. More observations for each source become available as the mission
evolves, thus improving the overall quality of the photometric outputs.

This method works because sources are observed under many different instrument conditions, each with its own calibration.  As long as there is a good mixing/overlap of all
the sources and CUs\footnote{In order to have good mixing, more than 50\% of the sources need to contribute to two or more CUs.} the iteration scheme will converge quickly.
If, for instance, there is no mixing of sources between two sets of CUs, the process can converge to two different photometric systems. This is not the case with {\Gaia} in
general, but there are some conditions that cause poor mixing between some CUs, where the convergence would be extremely slow.  In this case, an additional {\it link
calibration} is introduced to speed up the convergence (see Sect.~\ref{sec:internal}).

The goal is to reach precision at the S/N level for the faint sources while letting the precision for the bright sources be dominated by the calibration errors. Our
internal requirement is a calibration threshold of 1 mmag level per observation. The absolute calibration precision is constrained by the precision and accuracy of the
SPSS (see Sect.~\ref{sec:external}), and ultimately to the accuracy of Vega's spectra calibration, at a level of 1\%.

For many scientific purposes (stellar variability, photometric microlensing, exoplanet transits, morphology of solar system objects, etc.), the precision of the internal
fluxes of a given source is much more relevant than the precision of the absolute fluxes. For other purposes (galactic structure, clustering properties of galaxies, etc.),
the homogeneity of the calibration over large areas of sky is crucial. Again, what matters is the stability of the photometric instrument rather than the absolute fluxes.

For wide-field imaging surveys completed using on-ground pointing telescopes, the internal calibration usually delivers precision of a few percent. In fact, most
high-precision optical surveys with ground-based telescopes achieve better than 2\% precision. The most precise calibration accuracy is obtained from the SDSS-SNLS
supernova analysis (less than 1\%; \citealt{SDSSSNLS}). The main obstacle to reaching precise internal calibration well below the 1\% level is making the photometric
calibration consistent over large areas of the sky. {\Gaia} has the advantage of being in space (without the distortion of the Earth's atmosphere) and of being a scanning
satellite with two telescopes. Therefore, a given source is observed repeatedly over the duration of the mission (an average of 70 times in 5 years) following great circles
with different orientations on the sky. This naturally links the measurements in a given area with all other areas of the sky throughout the mission, thus providing the
natural mixing/overlap to ensure that all measures refer to the same internal/mean instrument.

For {\GDR1}, no standard source selection has been carried out for the internal calibration.  Owing to the complexity of the small scale calibration model (SS, see
Sect.~\ref{sec:internal}) and the limited number of observations available after just one year of the mission, it was necessary to maximise the number of standards
available for calibration by including all sources. Removal of variable sources from the calibrations is performed by iterating the calibration solutions with appropriate
outlier rejection. Although all the ingredients of the calibration pipeline have not been activated in this first release (see Sect.~\ref{sec:internal} and
\citealt{PhotValidation}), the quality of the internal calibration with the model explained here is at the millimagnitude level.

\section{Instrumental constraints}
\label{sec:instrument}

We summarise here for the sake of completeness some basic aspects of the
{\Gaia} instrumentation and data acquisition relevant to the photometric calibration. 
\cite{jordi2010} can also be consulted for more information about the
{\Gaia} photometric instrument. \cite{GaiaCCD} analyse the behaviour of the CCDs in the period relevant to {\GDR1}.

\begin{figure*}[t!]
 \centering
 \includegraphics[width=0.9\linewidth]{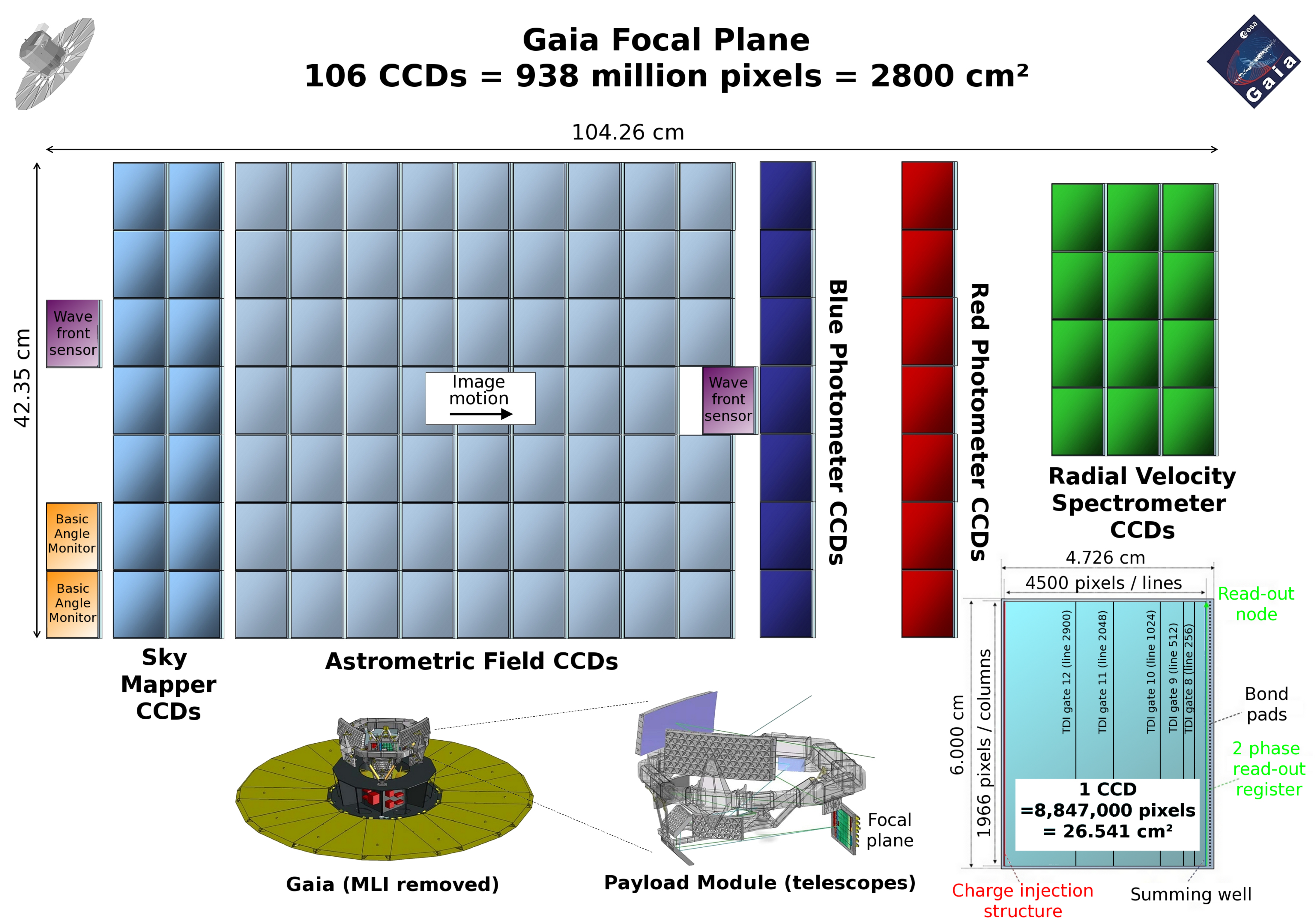}
  \caption{{\Gaia} focal plane. The viewing directions of both telescopes are superimposed onto this common focal plane which features 7 CCD rows, 17 CCD strips, and 106
large-format CCDs, each with 4500 TDI lines, 1966 pixel columns, and pixels of size 10 $\mu$m along-scan by 30 $\mu$m across-scan (59 mas $\times$ 177 mas). Source images
cross the focal plane from left to right.  For the photometric calibration explained here, only CCDs in the Astrometric Field and Blue and Red Photometers are relevant.
Picture courtesy of ESA - A. Short.
 \label{fig:focalplane}
}
\end{figure*}

{\Gaia} is a complex instrument (see overview of the focal plane in Fig.~\ref{fig:focalplane}), with 106 large-format CCDs (each with 4500 TDI\footnote{{\Gaia} CCDs are
operated in time delay integration (TDI) mode to collect charge from the sources at the same rate as they are transiting the CCD.} lines and 1966 columns, giving a total of
about one billion pixels) arranged in 7 rows (distributed vertically in Fig.~\ref{fig:focalplane}) and 17 strips (distributed horizontally in Fig.~\ref{fig:focalplane}).
The size of the focal plane is $104.26\times42.35$~cm$^2$, the largest sent to space to date. For more details see \cite{GaiaMission}.

While {\Gaia} scans the sky with its two telescopes, the images of the observed objects for both FoVs transit the focal plane along a given row, which are registered at the
different CCD strips. We can define a direction along (AL) and across (AC) the scan in the focal plane. Different components of the focal plane array are dedicated to
different purposes. The first two CCD strips (called Sky Mappers, SM) are used to detect the presence of any incoming point-like source in the FoVs. Each SM CCD strip sees
only one FoV, thanks to the use of appropriate masks. In each of the nine CCD strips in the astrometric field (AF), an image is recorded from which the flux in white light
photometry ($G$ passband, 330--1050~nm) and the centroid position, which provide the astrometry, are obtained.  Astrometric field fluxes are provided to the photometric
calibration processing once extracted by the Initial Data Treatment, \citep[IDT,][]{IdtRef}.  After transiting the AF field, the light is dispersed by two prisms in the AL
direction thus creating two low-resolution spectra:  one in the blue domain (BP, 330--680 nm) and \sout{the other} one in the red domain (RP, 640--1050~nm)\footnote{The
resolution for the BP and RP instruments is $30\lesssim R_{\rm BP}\lesssim 100$ and $60\lesssim R_{\rm RP}\lesssim 90$, respectively.}. The blue and red spectra are used to
classify and parameterise the observed sources and to account for chromaticity effects in AF centroiding measurements. The integrated flux of BP and RP spectra yield
{\BP} and {\RP} magnitudes as two broad passbands.  Finally, the radial velocity spectrometer (RVS) collects high-resolution spectra ($R\sim 11500$) to derive
radial velocities for the brightest objects ($G\lesssim 16$) and physical parameters for $G_{\rm RVS}\lesssim 12.5$~mag by using the spectral region around the CaII triplet
(845--872~nm).

Observations are acquired in different instrumental configurations (calculated in CUs) that need to be calibrated separately. Calibration of each CU is only possible when
enough observations have been collected. This implies that CUs covering a larger portion of the focal plane, a whole CCD for instance, will collect enough observations in a
shorter time and can be calibrated more often than those covering smaller portions (groups of CCD pixel columns, for instance). The calibration and linking of CUs to get a
unique photometric system is discussed in Sect.~\ref{sec:internal}.

We now describe the components defining the different CUs (see Table~\ref{tab:NCU} to see the CUs used for {\GDR1}):

\begin{description}

\item[\textbf{FoV:}] As previously stated, {\Gaia} has two telescopes. Although the CCDs are the same for 
both FoVs, the lightpath goes through different optical elements. Therefore, the two FoVs must be
calibrated separately
(see Sect.~\ref{sec:internal}).

\item[\textbf{Gates:}] {\Gaia} covers a very wide range of magnitudes ($3.0<G<20.7$~mag). To measure the magnitudes, {\Gaia} CCDs can be configured to integrate only the
charges collected over a fraction of the CCD, thus effectively reducing the exposure time.  The different configurations are called gate settings. The size of the section
of the CCD where integration takes place is different for different magnitude ranges and is chosen to minimise saturation\footnote{The magnitude ranges for each gate are
defined per CCD and stitch block.}. Observations taken with different gate configurations will be characterised by different integrated sensitivity of the pixels interested
by the transit. For this reason, observations obtained with different gates need to be calibrated separately.

\item[\textbf{Window class (WC):}]  

A small window of pixels around the central position of the object is defined on board in order to reduce the amount of telemetry and readout noise. Only this small portion
of the image is transmitted to the ground segment. To further reduce the readout noise and telemetry volume, the individual pixels inside the window are only completely
downloaded for brighter sources. For the fainter ones, the pixels inside the window are co-added electronically during readout and summed in the AC direction. The resulting
binned pixels are called `samples'. This means that we get 2D images for bright sources and 1D images for fainter ones. Also two different sizes of windows are set for
different magnitude intervals in the 1D regime. Table~\ref{tab:wc} summarises the types and magnitude ranges for the different WC defined.

\begin{table}
\begin{center}
\caption{Different types of windows defined depending on the magnitude of the source.\label{tab:wc}}
\begin{tabular}{cccc}
\hline
Window Class &    Type  & AF  & BP/RP \\
\hline
     WC0     &     2D	&  $G\leq13$	&  $G\leq11.5$    \\
     WC1     &  Long 1D & $13<G\leq 16$ & $11.5<G\leq 16$ \\
     WC2     & Short 1D & $G>16$	& $G>16$	\\
\hline          
\end{tabular}
\end{center}
\end{table}

\item[\textbf{AC position (row \& column):}] Observations obtained at different AC positions are affected by different responses (Figs.~\ref{fig:flatfield} and
\ref{fig:CRNU}) and different PSFs (see Figs.~\ref{fig:PSFfocalplane} and \ref{fig:PSF_FWHM}). These AC variations can be divided into two components: one causing a smooth
behaviour covering the whole focal plane (due to mirrors, passbands in BP/RP instrument, etc) and a second component causing a very different pixel-to-pixel response which
produces very different observing conditions from one column to the next.  Two of these effects causing strong differences in neighbouring pixels are the dead pixels and
the stitch block boundaries resulting from the CCD manufacturing process. The stitch blocks are 250 pixel columns wide, except for the two outermost blocks; the exact block
boundaries are at $\mu_b = 13.5$, 121.5, 371.5, \ldots , 1621.5, 1871.5, 1979.5. Figure~\ref{fig:CRNU} shows the effect of a dead pixel at column 1253 and of the stich
block around column 1370, especially visible in the 900~nm case (see \citealt{GaiaMission} for more information). The first component, can be handled by the large scale
(LS) calibrations by adding a smooth function of AC coordinate. The pixel-to-pixel variations are handled by the small scale (SS) calibration which is solved over small
groups of pixel columns (four columns were used in {\GDR1} photometric calibration) per gate and WC (see Sect.~\ref{sec:internal}).

 \begin{figure}
  \begin{center}
   \includegraphics[width=0.16\textwidth,height=4cm]{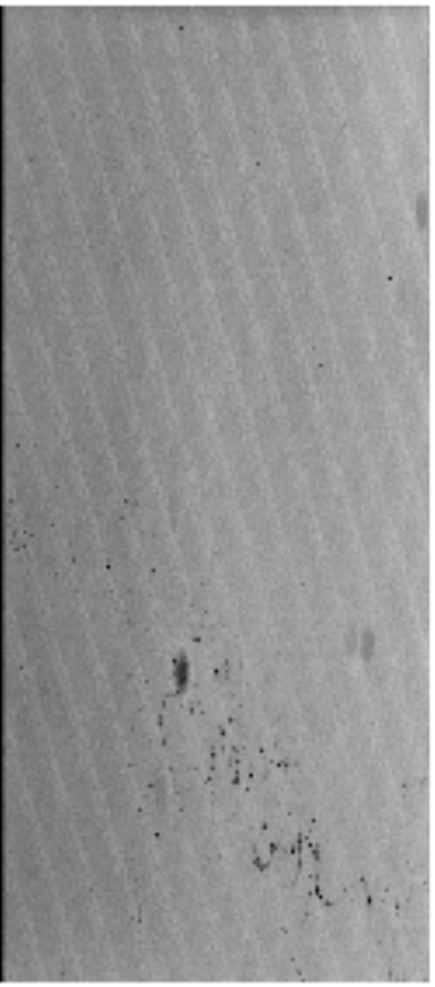}
   \includegraphics[width=0.16\textwidth,height=4cm]{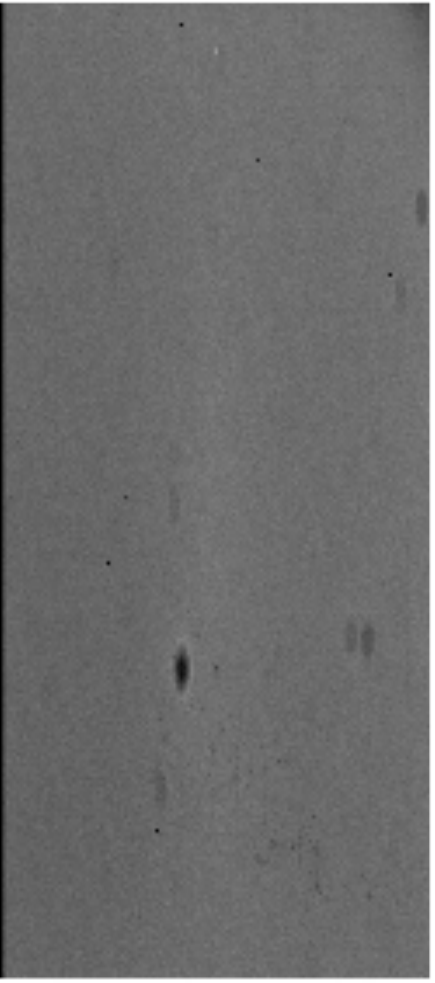}
   \includegraphics[width=0.16\textwidth,height=4cm]{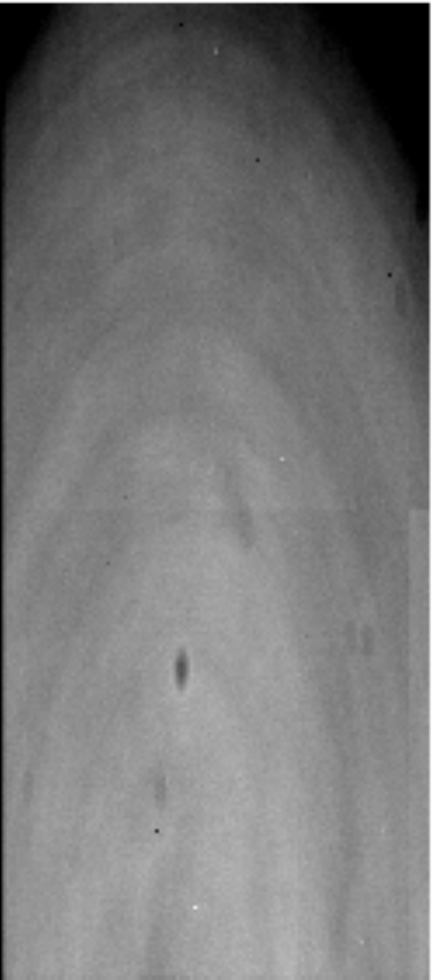}
   \end{center}   
	\caption{\small{Flatfield images obtained during pre-launch tests for 
one of the AF CCDs
for $\lambda=400$~nm, $550$~nm and $900$~nm (from left to right). CCD columns are aligned vertically (AL scan direction) in this figure.
All images are linearly scaled between the pixel values of 25\ 000 and 33\ 000 ADU.
Picture courtesy of
ESA - e2V. }
\label{fig:flatfield}
}    
\end{figure}

  \begin{figure}
  \begin{center}
   \includegraphics[width=0.5\textwidth]{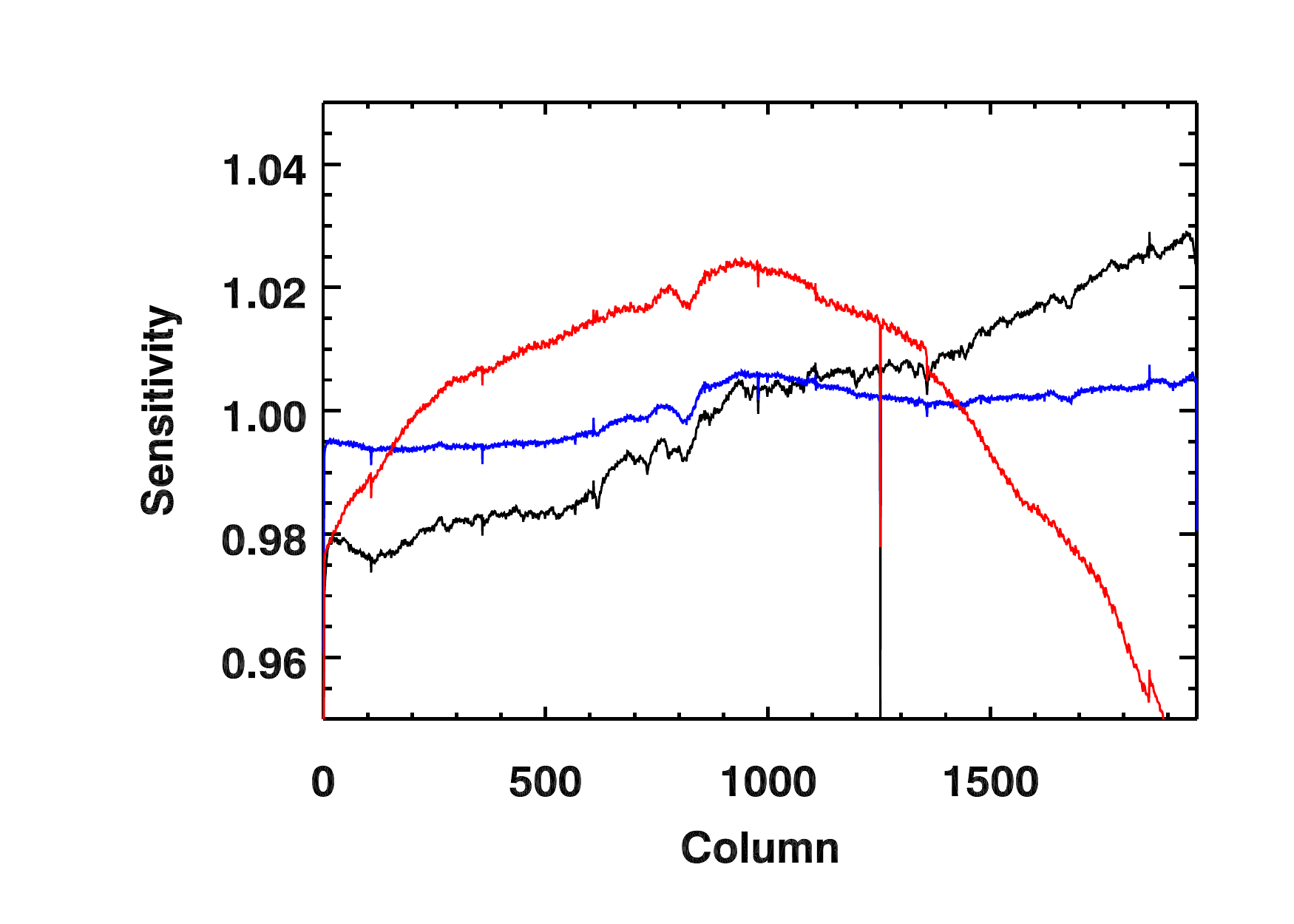}
   \end{center}   
	\caption{\small{Column-response non uniformity (CRNU) measured for the same {\Gaia} CCD plotted in Fig.~\ref{fig:flatfield}
for  $\lambda=400$~nm (blue line), 550 nm (black line) and 900 nm (red line). 
At column 1253 we can see the effect of a dead pixel.
The discontinuity around column 1350 is due to the stitch blocks.
Data courtesy of
ESA - e2V.}
\label{fig:CRNU}
}    
\end{figure}

 \begin{figure}
  \begin{center}
   \includegraphics[width=0.49\textwidth]{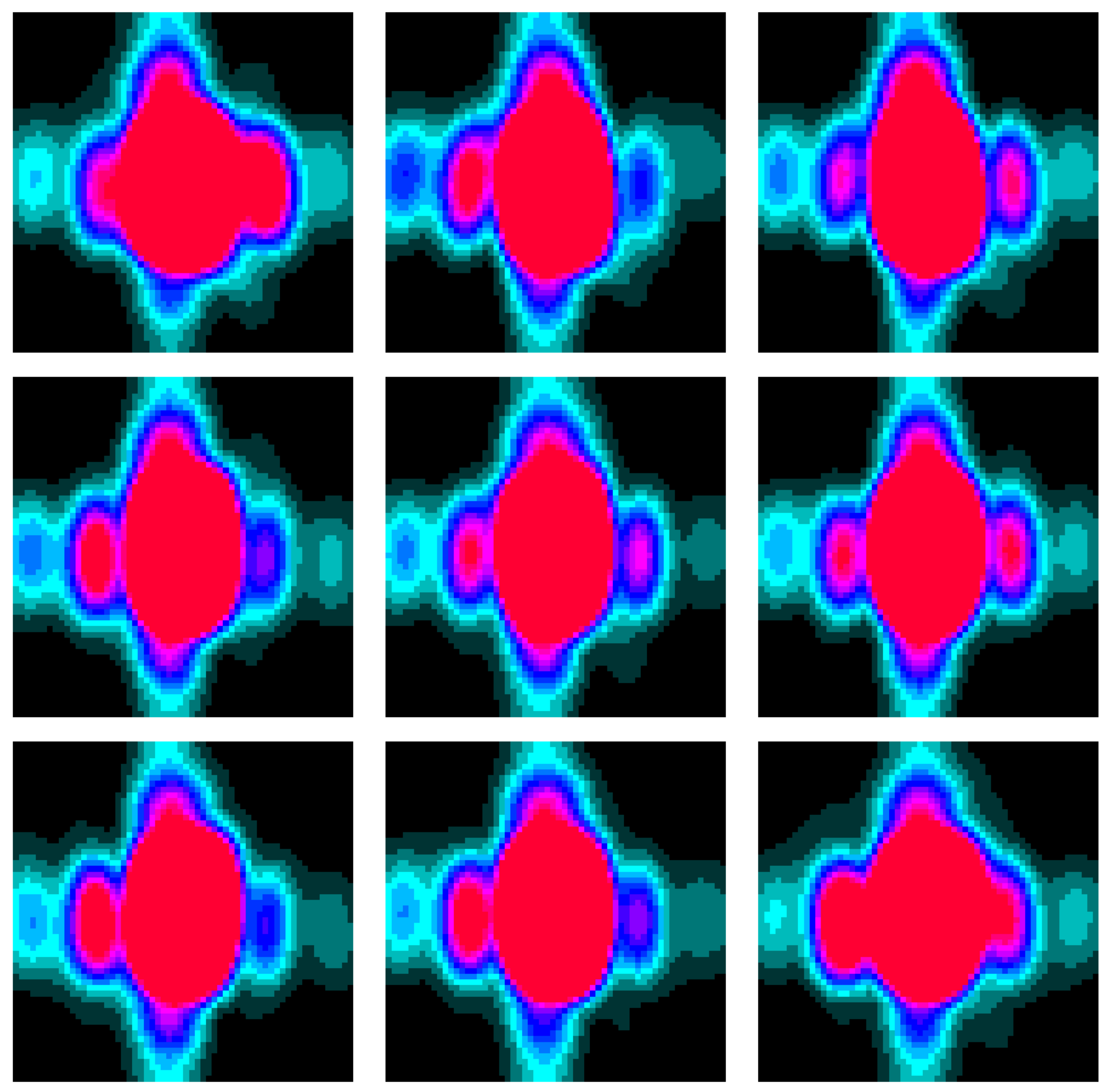}
   \end{center}   
	\caption{\small{Pre-launch simulated effective PSFs (using GIBIS simulator, see \citealt{GIBIS}) of a red source ($V-I=-0.5$~mag) and nominal (mean) AC scan motion 
for AF S1R1, S5R1, S9R1, S1R4, S5R4, S8R4, S1R7, S5R7 and S9R7 of telescope 2 (starting in the top left corner), meaning S the strip and R the row position of the CCD in
the focal plane respectively. These PSF variations with the focal plane position are also dependent on the colour of the source and the FoV.}
\label{fig:PSFfocalplane}
}    
\end{figure}

 \begin{figure}
  \begin{center}
   \includegraphics[width=1.0\columnwidth]{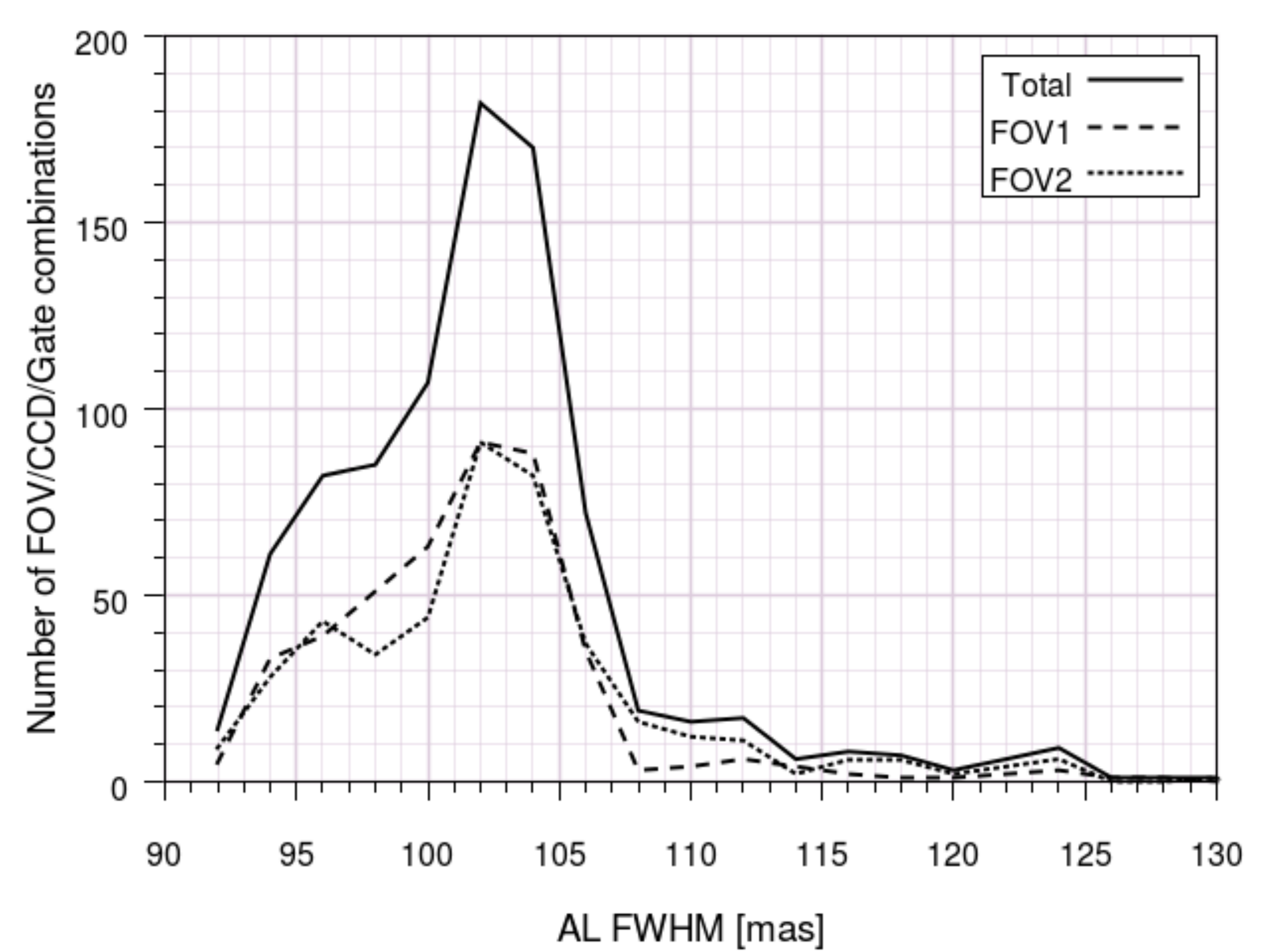}
   \end{center}   
	\caption{Histogram of the FWHM in the AL direction for all CUs in AF at a mean colour.
	The solid black line shows the total; the dashed and dotted lines show 
        the preceding and following FoV
        respectively. The median FWHM is 103 mas (1.75 pixels). Figure extracted from \cite{IdtRef}.
	\label{fig:PSF_FWHM} 
}    
\end{figure}


\item[\textbf{Time:}] Ageing of the instrument (radiation effects, see \citealt{GaiaCCD}) and other time dependent effects 
(e.g. contamination and straylight effects) cause the instrument response to 
 vary significantly throughout the mission. For example, for the instruments on board the HST, \cite{bohlin2007} estimates a loss of sensitivity with time of 0.2--0.3\% per
year for intermediate wavelengths and 0.8\% per year for red wavelengths. For this reason only observations close in time can be considered for a given calibration. The
instrumental time dependencies present in {\GDR1} are dominated by the contamination effects (see below).  Different CUs must then be defined in different time ranges,
paying particular attention to times at which abrupt changes are expected in the calibrations (e.g. decontamination events).

\end{description}

For each CU, several instrumental effects must be accounted for(see also Fig.~\ref{Fig:IterationPlot}):

\begin{description}

\item[\textbf{Overall response:}] In the final observation we cannot distinguish between the contributions of sensitivity variations in the detectors from transmissivity
variations due to the mirrors and contamination issues. Then, an overall response is evaluated. The response varies across the 106 CCDs.  Figure~\ref{fig:flatfield} shows
three pre-launch flatfield images for three different wavelengths. As can be seen, the sensitivity at red wavelengths strongly decreases at the corners of the CCD (having a
3--4\% of variation in different regions). As the image of the star is transiting all CCD TDI lines while {\Gaia} spins, it is necessary to know the accumulated effect of
all pixels in a column. This is termed the column response non-uniformity, or CRNU (Fig.~\ref{fig:CRNU}). The measured fluxes in each observation suffer from sensitivity
variations which modify the size of the signal in a colour-dependent way. At central wavelengths, 550~nm, this accumulated effect is only $\pm 1$\%, but at extreme
wavelengths, 400 and 900~nm, it is of the order of $\pm 4$\%. The calibration of this effect is dealt with in Sect.~\ref{sec:internal}.  The quantum efficiency (QE) also
changes from CCD to CCD. Pre-launch estimations evaluated a difference of $\pm 1$\% at 550~nm, but increasing to 3 and 6\% for more extreme wavelengths (900 and 400~nm
respectively). CCD columns with dead pixels (see column at 1253 in Fig.~\ref{fig:CRNU} as an example) can be masked during IDT \citep{IdtRef}. Also, different CUs are
considered for the different stitch blocks (see Sect.~\ref{sec:instrument} and \ref{sec:colour}).

\item[\textbf{Aperture effect:}] When defining the small window around a source, part of the flux is cut off. In the case of 2D windows, a PSF model is fitted to the
samples in the window\footnote{For {\GDR1} this PSF model is quite simple and two 1D profiles of the PSF (called Line Spread Function, LSF), one in each
direction, AL \& AC, are multiplied \citep{IdtRef}.}. In this case, no aperture correction is needed. However, for 1D windows, only the AL flux of the observation can be
fitted. This does not account for the part of the flux outside of the window in the AC direction. The size of the effect depends on the AC centring offset of the source in
the window, the AC motion\footnote{Owing to the nominal precession of the spin axis, the AC position can change substantially in the focal plane during the observation (up
to 4.5 pixels in a CCD transit, see \citealt{IdtRef}). This AC motion produces an AC smearing and increases the amount of flux loss.} and the colour of the
source\footnote{The dependence on colour is a consequence of the PSF dependence on colour.}.

Calibration of the aperture effect is omitted from {\GDR1} owing to the lack of available astrometric information at the time of photometric processing needed to calculate
the centring offsets at the required accuracy. Future releases will include the calibration of this effect. Using pre-launch {\Gaia} PSFs and realistic distributions for
centring offsets, colours and AC velocities, it was estimated that this effect adds a 5--10 mmag scatter to the overall photometric errors of individual 1D flux
measurements. Figure~\ref{fig:loss} shows the size of the aperture effect for a given CCD as a function of colour, AC motion, and centring offset. Considering all CCDs
together, the effect of the centring offset on the amount of flux loss is lower than 0.9\%. The effect of the AC motion is lower than 0.4\%. The colour effect, on the other
hand, produces the largest variations (0.4--1.1\%).

\begin{figure}
  \begin{center}
    \includegraphics[width=0.5\textwidth]{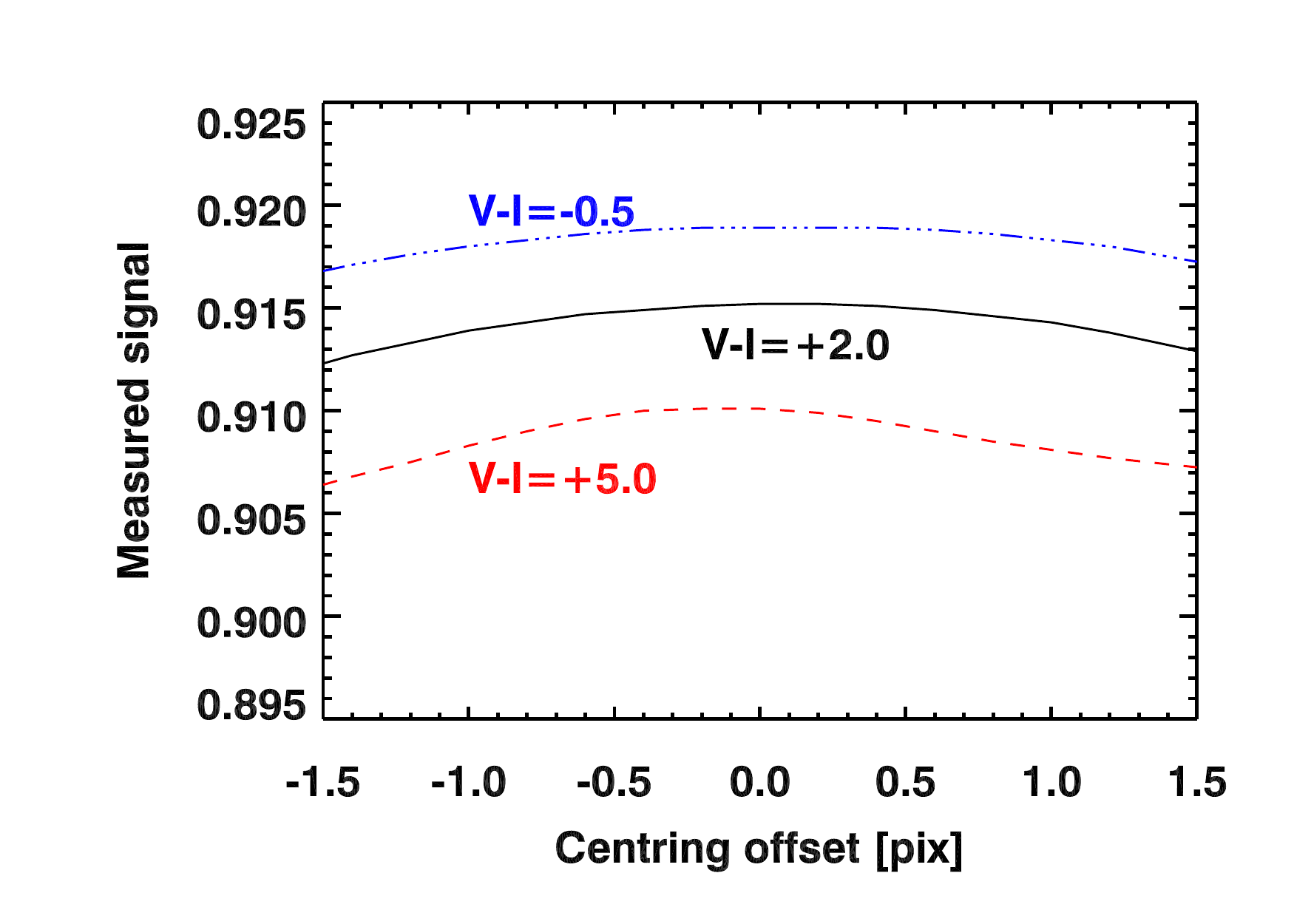}
   \includegraphics[width=0.5\textwidth]{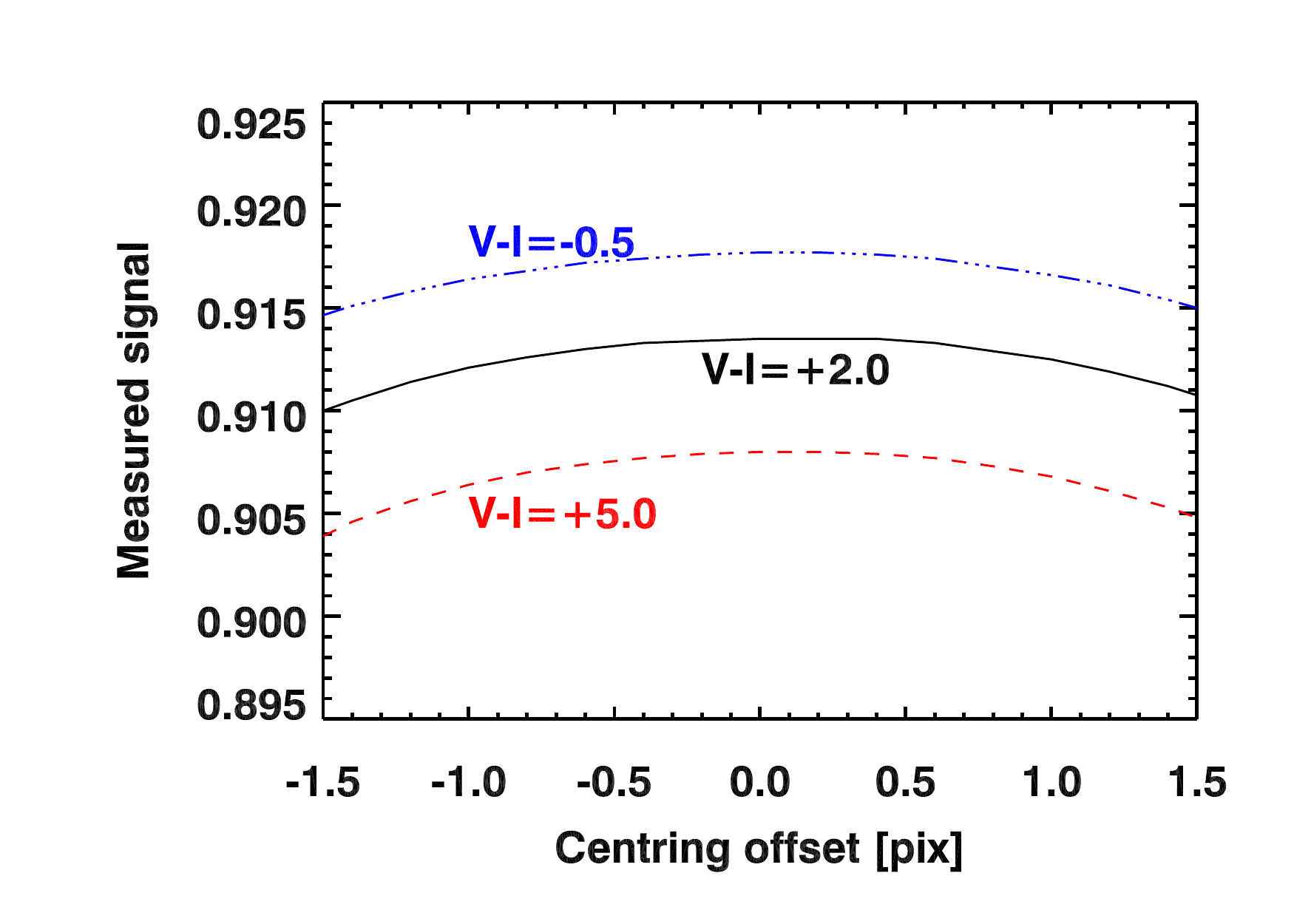}
  \end{center}  
	\caption{\small{The fraction of measured flux in the CCD for strip 9, row 1 and FoV 2 in AF is plotted
	as a function of the centering offset with no AC motion
(top) and maximal AC motion (bottom). For assumed maximal centering offsets of about one pixel the dependence on the 
colour $V-I$ of the 
source is the dominating effect for AC flux loss with variations up to 1.1 percent.}
\label{fig:loss}    
}
\end{figure}

\item[\textbf{Contamination:}] A colour dependent degradation of the overall response with time due to the presence of water ice contaminant deposited on the mirror
surfaces of the telescopes and focal plane was detected during the commissioning phase. This effect is mitigated by decontamination events, heating up the mirrors and the
focal plane when a threshold response degradation is reached. These decontamination events can be accounted for by allowing for abrupt changes in the calibration parameters
(see \citealt{PhotProcessing}).  Contamination effects can change the measured magnitudes in {\GDR1} by 0.2-0.3~mag, depending on the colour of the source and the level of
contamination \citep{GaiaMission}.

\item[\textbf{Background \& straylight:}] Background subtraction is part of pre-processing: in IDT for $G$ \citep{IdtRef} and in PhotPipe for BP/RP (see
Sect.~\ref{sec:colour}). The background has contributions from non-resolved stars, zodiacal light, etc.  Also, unwanted straylight reaches the focal plane leading to an
increased average background level and to significant variations in the background depending on the spin phase.

\item[\textbf{Geometry:}] The position and orientation of each CCD needs to be calibrated. In addition, the focal length of each telescope is slightly different. This
influences the propagation of the position of the window to be read at every CCD in a given transit. This causes an imperfect centring of the observed source inside the
window thus producing extra flux losses that can be accounted for during the calibration process. In BP/RP observations, these shifts in the AL direction can also affect
the wavelength calibration and need to be evaluated before extracting any colour information (Sect.~\ref{sec:colour}).

\item[\textbf{Other effects:}] Other instrumental effects (such as gain, bias, dark current) are accounted for in the preprocessing. The effects of charge transfer
inefficiency (CTI) during image and serial register readout are due to radiation damage and are not accounted for in {\GDR1}. These were evaluated to be small at the
beginning of the mission \citep{GaiaCCD}, and are expected to be more important as the mission evolves. Ideally, the signal level increases linearly with exposure time
until image area full well capacity (FWC) is reached where it becomes constant. If the anti-blooming structure shows a soft turn-on, even for a single column, the signal
will deviate from the linear response before FWC is reached and slopes gradually towards the column saturation level for increasing exposure time (see
Fig.~\ref{fig:nonlinear}). Saturated samples (see Fig.~\ref{fig:satlevels} for the saturation levels for one of the AF CCDs) can be masked during the fitting of the
LSFs/PSFs (although this has not yet been applied in {\GDR1}; see \citealt{IdtRef}).  Gates are chosen to limit the frequency of saturated observations to a maximum of 5\%.

\begin{figure}
  \begin{center}
    \includegraphics[width=0.5\textwidth]{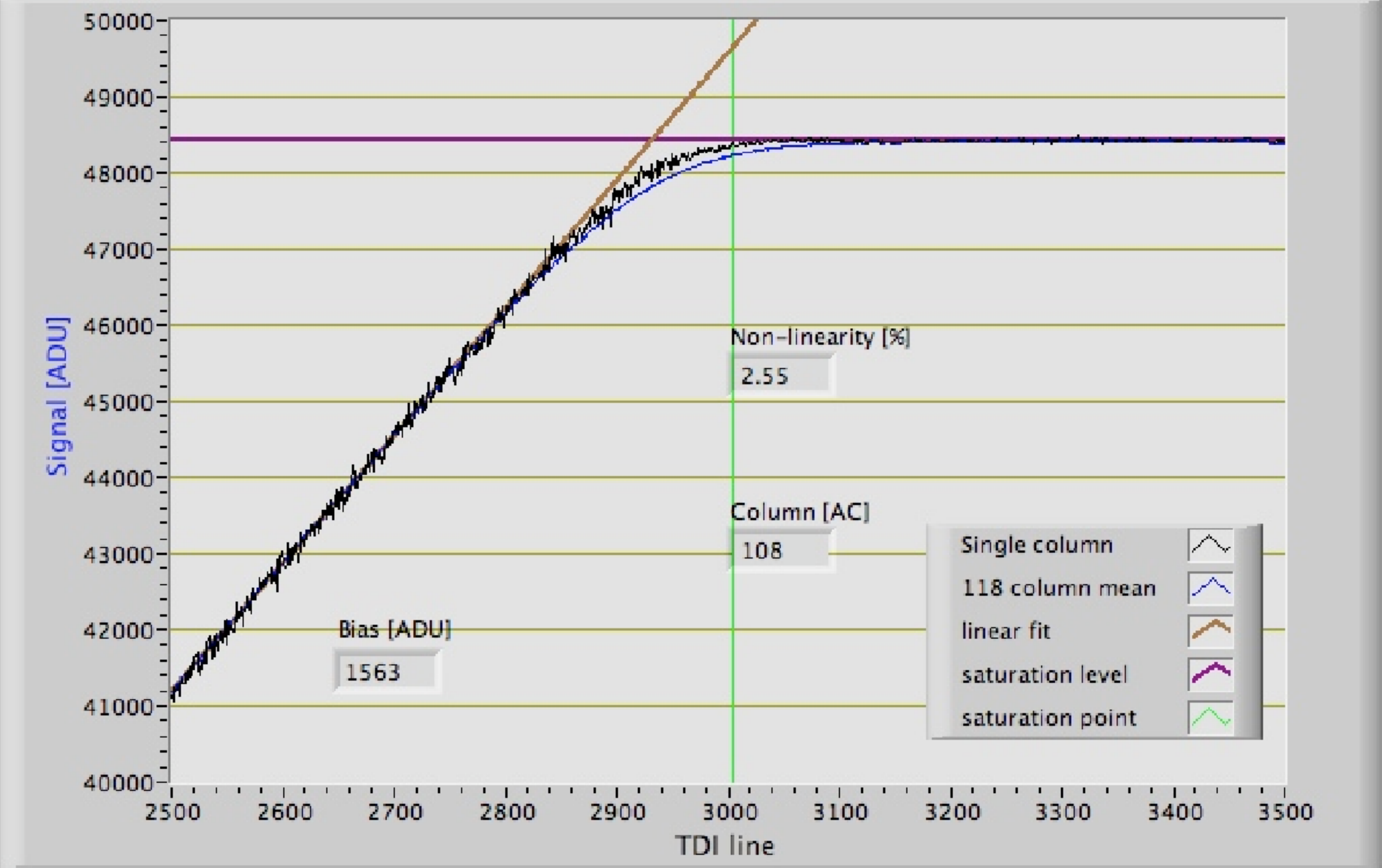}
  \end{center}  
	\caption{\small{Single column response to linear increase of illumination level per TDI line.  The black lines represent the data points of single pixel samples
from column 108. The blue curve represents the mean response of all columns analysed.  The vertical green lines mark the saturation point for the column, defined as the TDI
line, at which the signal enters the $3\sigma$ zone around the mean saturation level.  The mean value of non-linearity (deviation from expected linear response) for all 118
columns used in this test is $2.82\pm0.48$\%. The plotted column has a non-linearity equal to 2.55\%.  Plot courtesy of Ralph Kohley, Astrium and MSSL.
 }
\label{fig:nonlinear}    
}
\end{figure}

\begin{figure}
  \begin{center}
    \includegraphics[width=0.5\textwidth]{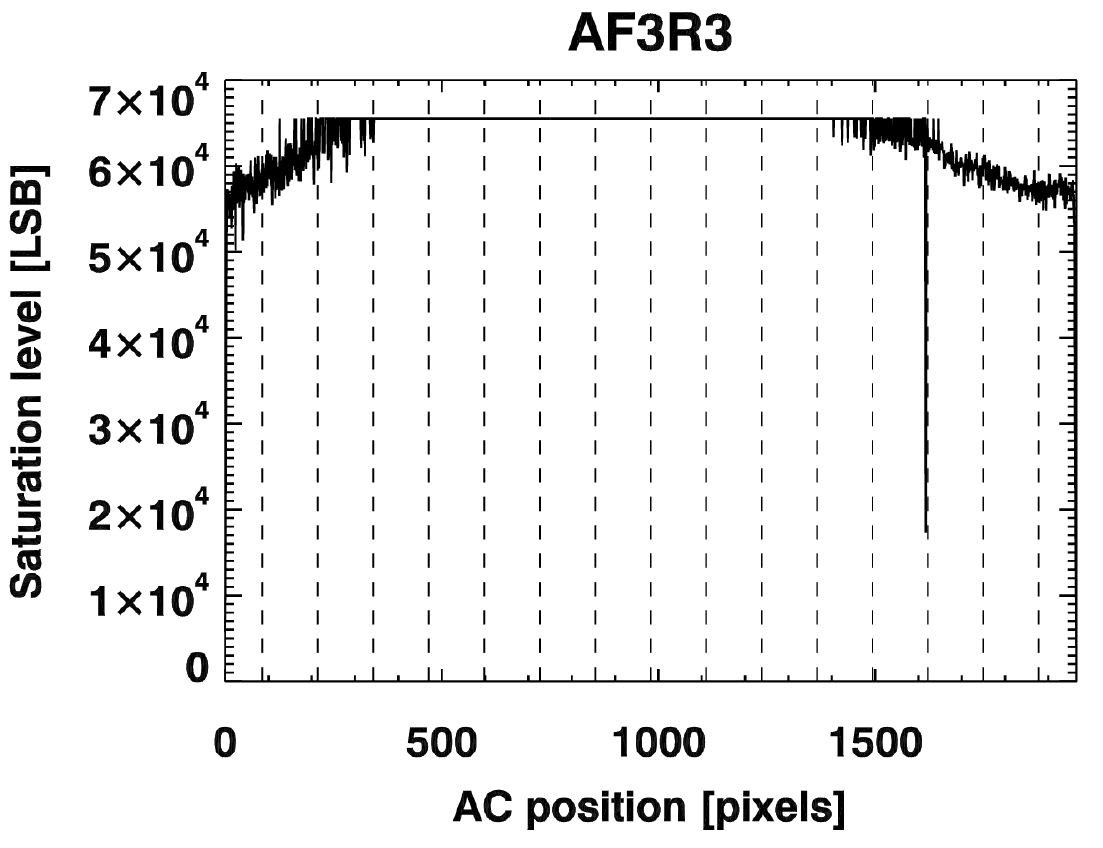}
  \end{center}  
	\caption{\small{Saturation levels (measured in LSB, Least Significant Bit) for the AF CCD of strip 3 and row 3. The dashed vertical lines show the borders of the 16
stitch blocks for which magnitude intervals for the activation of the gates are determined. Most of the CCDs show saturation level variations of less than 10000 LSB. Large
areas of the CCD are dominated by analogue-to-digital conversion saturation and only at both edges image area saturation is dominating.}
\label{fig:satlevels}    
}
\end{figure}

\end{description}

\section{Internal calibration}
\label{sec:internal}

In this section we describe the formulation of the internal calibration model (see Fig.~\ref{Fig:IterationPlot}). This model is applied to all three {\Gaia}
passbands ($G$, {\BP} and {\RP}), although in {\GDR1} only $G$ photometry is released.

The main task of the internal calibration is to determine the variation of the overall response of the photometric instrument
over the mission and between different CUs (see Sect.~\ref{sec:instrument}). The $G$ flux calibration model describes these
variations for the entire instrument including the mirror assembly and the CCDs.
For the BP and RP channels, the prisms also constrain the
total amount of flux measured. 

As explained in Sect.~\ref{sec:principles}, a requirement of the internal instrument calibration is that it should have enough
observations of the standard sources in the considered CUs. The CUs must be defined in order to ensure this requirement and
the calibration can only be computed when enough observations and sufficient overlap/mixing are obtained for that CU. For this reason, the calibration model
was split into 
LS and SS components.

The SS calibration models the variations of the sensitivity at the CCD column level, whereas the LS calibration accounts for variations in the mean
response at the CCD level and for every FoV. 
It naturally follows that more observations will be available for the calibration of a LS CU than for
a SS one and this will allow the calibration of LS effects on a shorter timescale (of the order of one day).

The following equation describes the relation between the internal reference flux, $\overline{I_s}$,
of a source $s$ and its measured flux, $I_{skll^{\prime}}$,
in a given observation $k$ obtained in a given LS unit, $l$, and a given SS unit, $l^{\prime}$,

\begin{equation}
\label{eq:Gxpmodel}
\frac{I_{skll^{\prime}}}{\overline{I_s}}=\ensuremath{\text{LS}}_{skl} \cdot \ensuremath{\text{SS}}_{skl^{\prime}}
\end{equation}

\noindent where $\ensuremath{\text{LS}}_{skl} \cdot \ensuremath{\text{SS}}_{skl^{\prime}}$ is the product of the LS 
and SS calibration models applicable to the observation. 
Both $\overline{I_s}$ and $I_{skll^{\prime}}$ are expressed in photoelectrons per second.

In the following equation, we describe $\ensuremath{\text{LS}}_{skl}$ as an $R$th degree 
polynomial function depending on $M$ ``colour terms'', $C_{sm}$. In addition to
this colour term, a $J$th degree polynomial dependency with the AC position, $\mu_k$,  of the observation is introduced to account for smooth variations across the focal plane.

\begin{equation}
\label{eq:GxpmodelLS}
\ensuremath{\text{LS}}_{skl}
=
\sum_{r=1}^{R} \sum_{m=1}^{M}  A_{rml}  \cdot (C_{sm})^r  + 
\sum_{j=0}^{J} B_{jl} \cdot (\mu_k)^j
\end{equation}

Analogously, $\ensuremath{\text{SS}}_{skl^{\prime}}$ is described by an $R'$th degree polynomial
depending on $M^{\prime}$ colour terms:

\begin{equation}
\label{eq:GxpmodelSS}
\ensuremath{\text{SS}}_{skl^{\prime}}
=
\sum_{r=0}^{R^{\prime}} \sum_{m=1}^{M^{\prime}} a_{rml^{\prime}} \cdot (C_{sm})^{r}
\end{equation}

We note that the SS model for {\GDR1} only implements the zero point ($R^{\prime}=0, M^{\prime}=1$).

The coefficients $A_{rml}$ and $B_{jl}$ are the LS (CCD \& FoV) and $a_{rml^{\prime}}$ the SS (group of columns) instrumental
coefficients, respectively, valid for the \textbf\sout{CU associated with} $l$ and $l^{\prime}$ CUs corresponding to observation $k$.

The subindexes $l$ and $l^{\prime}$ are in fact a simplification of the different subindexes defining a given CU. 
For {\GDR1} we defined a LS CU for every combination of CCD row, CCD strip, Gate/WC, FoV, and unit of time, and a SS CU
for every CCD row, CCD strip, Gate/WC, 4 column bin, and unit of time. The time ranges used for {\GDR1} photometric 
calibration is 1 day for the LS and 14 months for the SS.
Table~\ref{tab:NCU} lists the number of CUs used for {\GDR1} photometric calibration. Table~\ref{tab:Nobs} 
gives the approximate number of observations available in {\GDR1} processing for each CU for LS and SS calibrations.

\begin{table}
\begin{center}
\caption{Approximate number of observations per CU for LS ($N_{\rm obs}^{\rm LS}$)
and SS ($N_{\rm obs}^{\rm SS}$)
calibrations used for {\GDR1}. The 
$t_{\rm exp}$ column shows the effective exposure time for the different gate configurations. The $G$ range column 
provides only approximate values.}
\tiny
\begin{tabular}{p{0.7cm}p{0.7cm}p{0.7cm}p{0.7cm}p{1.2cm}p{1.1cm}p{1.1cm}}
\hline
Instrum. & Window & Gate & $t_{\rm exp}$~(s) & $G$ range &  $N_{\rm obs}^{\rm LS}$ & $N_{\rm obs}^{\rm SS}$\\
\hline
AF       & WC0 & Gate04 & 0.02 & $G<8.5$    & 300	  & 400    \\
AF       & WC0 & Gate07 & 0.13 & 8.5--9.5   & 300	  & 450    \\
AF       & WC0 & Gate08 & 0.25 & 9.5--10.0  & 600	  & 1000   \\
AF       & WC0 & Gate09 & 0.50 & 10.0--11.0 & 700	  & 1100   \\
AF       & WC0 & Gate10 & 1.00 & 11.0--12.0 & 1900	  & 3000   \\  
AF       & WC0 & Gate11 & 2.01 & 12.0--12.2 & 1000	  & 1200   \\  
AF       & WC0 & Gate12 & 2.85 & 12.2--12.4 & 1800	     & 2500	   \\  
AF       & WC0 & None	& 4.41 & 12.4--13.0 & 12\ 000	  & 23\ 000	\\   
AF       & WC1 & None	& 4.41 & 13.0--16.0 & 150\ 000    & 290\ 000	\\    
AF       & WC2 & None	& 4.41 &  $G>16.0$  & 2\ 200\ 000 & 3\ 600\ 000 \\   
\hline          
\end{tabular}
\label{tab:Nobs}
\end{center}
\end{table}

The $C_{sm}$ and $C_{sm'}$ terms in Eqs.~\ref{eq:GxpmodelLS} and \ref{eq:GxpmodelSS} account for the fact that
the overall response variations 
are wavelength dependent.
The colour information is retrieved from the {\Gaia} BP/RP observations.

The simplest way to get colour information from {\Gaia} data is to compute the ({\BP}$-${\RP}) colour. Another possibility is to integrate the BP/RP spectra in several
wavelength ranges. This approach is used in PhotPipe since this allows a more detailed characterisation of the SED than with a single {\BP}$-${\RP} colour.

The results of this synthetic photometry are called Spectral Shape Coefficients (hereafter SSC). The SSC-based colours 
and the way they are extracted from the observations are described in Sect.~\ref{sec:colour}.

\begin{figure}[t!]
 \centering
 \includegraphics[width=1.0\linewidth]{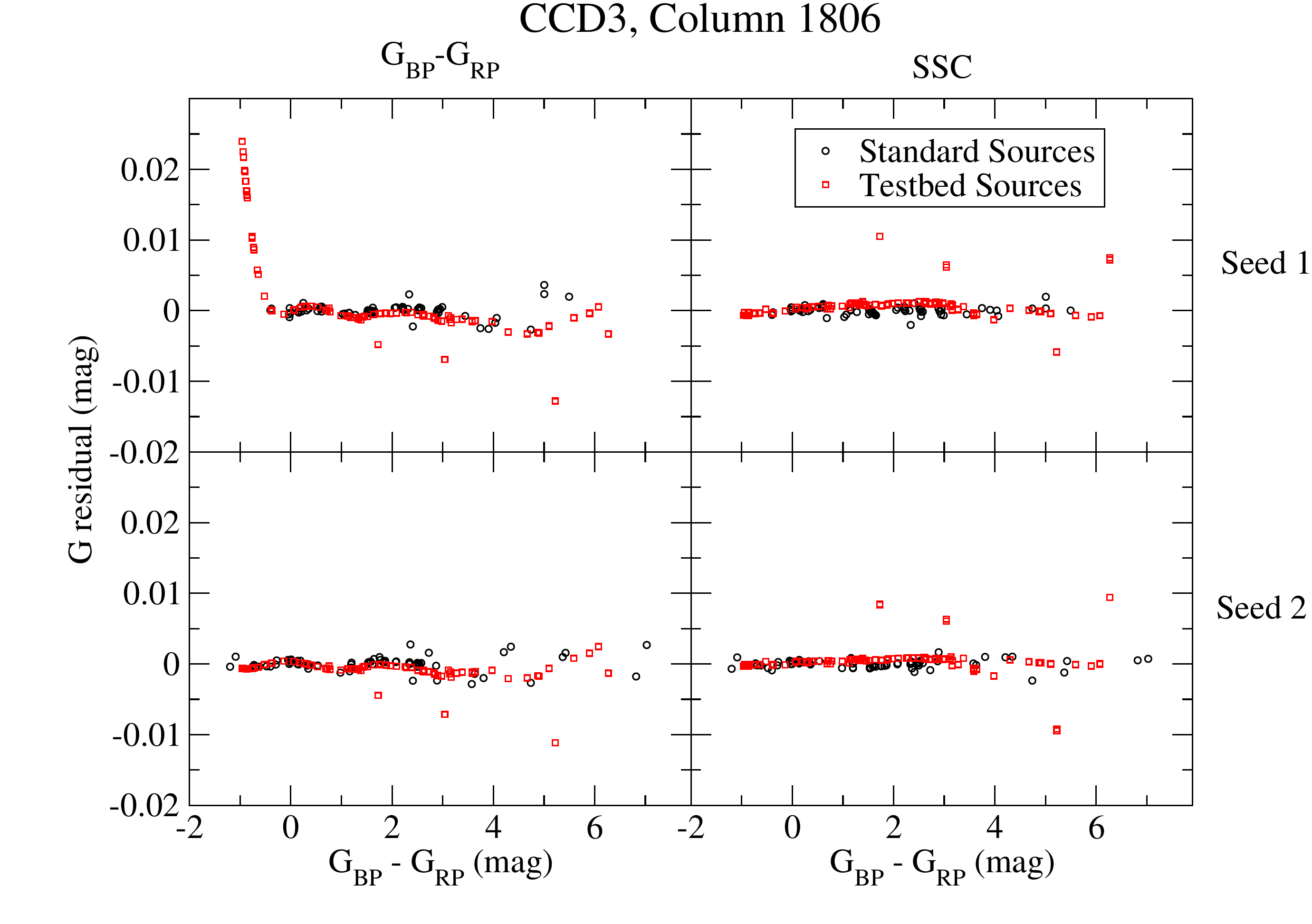}
  \caption{Dependency of the residuals in $G$ on column 1806 of CCD row 3 when using two different colour distributions of the standard sources (see text).
 \label{fig:resGseed}
}
\end{figure}

Using SSCs as the $C_{sm}$ and $C_{sm'}$ terms, we improve the calibration residuals and decrease the degree of the polynomial with respect to the single {\BP}$-${\RP}
colour alternative. The systematics for extreme colours and for outliers are similarly reduced. SSCs also make the standard source selection strategy less critical.
Figure~\ref{fig:resGseed} illustrates the importance of reducing the degree of the polynomial, and of being sure that the selection of standard sources is done with the
widest possible range of colours. We simulated the photometric calibration process with synthetic SED libraries (using the Gaia Object Generator, GOG; \citealt{gog}) using
the same set of standard sources, but with a different random distribution of their observations on the different columns in every CCD. Consequently, the colour
distribution of standard sources changes in every column.

In addition to simulations for the calibration sources (black circles in Fig.~\ref{fig:resGseed}), we also simulated test sources 
(red circles) to check their residuals. The test sample is composed
of a total of 100 sources of different types with respect to the calibration sources, including emission line stars, quasars and very cold 
stars. Reddening effects were also simulated.

Figure~\ref{fig:resGseed} shows the residuals in $G$ for one column (column 1806 in CCD row 3). The application of the calibration coefficients to the test sources in the
`seed 1' case (upper panels) provides larger residuals when using the single-colour approach ({\BP}$-${\RP})  than when using the SSC model. For the `seed 2' case (bottom
panels), the colour interval of the standard sources covers the same range as the test sources. None of the models has problems in this case, although SSC yields lower
scatter than {\BP}$-${\RP}. The result confirms expectation: the colour distribution of the reference sources is less critical when adopting a linear model in more than one
colour (SSCs in this case) than in the case of a higher degree model in a single {\BP}$-${\RP} colour.

We define four SSC passbands for each of BP and RP (see Sect.~\ref{sec:colour}). To decrease the correlations between the different SSCs, it was decided to reduce the
number of colour terms by summing the two central SSCs in both BP and RP. The decision to sum these values was made because the sensitivity in the central wavelengths is
rather similar, while it varies more at the extreme wavelengths (see Figs.~\ref{fig:flatfield} and \ref{fig:CRNU}).

In the case of the $G$ flux calibration, the SSC information coming from both BP and RP is relevant, as the $G$-band covers 
their entire wavelength range. 
This implies that for $G$, the LS flux calibration model has $M=6$ in Eq.~\ref{eq:GxpmodelLS} (three $C_{sm}$ terms for the BP SSCs and three more for RP SSCs) with a
linear dependency in colour ($R=1$).
The degree of the AC position polynomial adopted is $J=2$. 

Table~\ref{tab:ncoefs} summarises the number of parameters for every CU.

\begin{table}[!htbp]
  \caption{Number of coefficients used in {\GDR1} for Eqs.~\ref{eq:GxpmodelLS} and \ref{eq:GxpmodelSS} for a given CU ($l$ or $l'$).
  \label{tab:ncoefs}  
  }
  \begin{center}
    \begin{tabular}{cccc}
\hline
   &  Equation Limits    & $N_{\rm coefs}$ & Coefficients \\
\hline
   &                     &                 & $A_{11l}$, $A_{12l}$, $A_{13l}$, \\
LS & $R=1$, $M=6$, $J=2$ &       9         & $A_{14l}$, $A_{15l}$, $A_{16l}$,\\ 
   &                     &                 & $B_{0l}$, $B_{1l}$, $B_{2l}$ \\
\hline
SS & $R'=0$, $M'=1$       &       1         & $a_{01l'}$ \\
 \hline
    \end{tabular}
  \end{center}
\end{table}

The implementation of Eq.~\ref{eq:Gxpmodel} (see Fig.~\ref{Fig:IterationPlot}, Sect.~\ref{sec:principles} and \citealt{PhotProcessing} for more details) is an iterative
process in which the different elements of the equation are determined assuming the others as known. In the first iteration, $\overline{I_s}$, $C_{sm}$ and $C_{sm'}$ are
derived through the weighted mean (see Sect.~\ref{sec:update}) of all the observations of every standard source. The values of $A_{rml}$, $B_{jl}$ and $a_{r'm'l}$ have been
determined, and $C_{sm}$ is also updated. Once a first set of calibration parameters are available at the end of the first iteration, they are applied to the observed
fluxes. New weighted averages ($\overline{I_s}$) are thus computed to improve the catalogue of reference fluxes to be used in the subsequent iteration.

In the first iteration, $\ensuremath{\text{SS}}_{skl'}$ is assumed to be 1.
The residuals obtained from the application of the LS calibration are then fitted to derive the SS 
coefficients. In the calibration process performed for {\GDR1} no divergence problems appeared when iterating the LS and SS
processes, so no renormalisation of the instrumental coefficients was needed.


When setting up the reference fluxes ($\overline{I_s}$ in Eq.~\ref{eq:Gxpmodel}), during the initial set of iterations of the
LS calibrations and reference flux calculations, a sufficiently large number of sources ($>50\%$) entering two or more CUs
is needed to avoid  the derived reference fluxes converging to different photometric systems (see Sect.~\ref{sec:principles}). In
that case, each independent CU would form its own system from the average 
for that group. 
This is avoided because some sources are observed in different CUs, and that mixing naturally provides convergence to a single instrument. 
However, if the mixing is poor, convergence will be very slow and a separate calibration is desirable in order to speed up the
iteration process. 
In fact, this calibration acts as an additional constraint imposed at the start of the iterations to ensure the consistency of the
internal photometric system over all instrument configurations.

One case identified early on as having poor mixing was that between the different gating and WC configurations
(see Sect.~\ref{sec:instrument}). Determining which gate or WC is chosen for an observation is done
on board using the magnitude of the source as measured by the SM. If this determination is accurate, each source will only be observed
using one gating and WC configuration. Fortunately, for sources brighter than $G=12$~mag the precision of this
on-board measurement is low ($\sim 0.3$~mag).
However, for
$G=13$~mag, which is 
the transition between 1D and 2D observations, the accuracy is quite good ($0.01$~mag) and therefore the mixing is
poor. Additionally, between the 1D and 2D WCs there is the difference in flux due to the aperture correction (see
Sect.~\ref{sec:instrument}).

  \begin{figure}
  \begin{center}
   \includegraphics[width=0.4\textwidth]{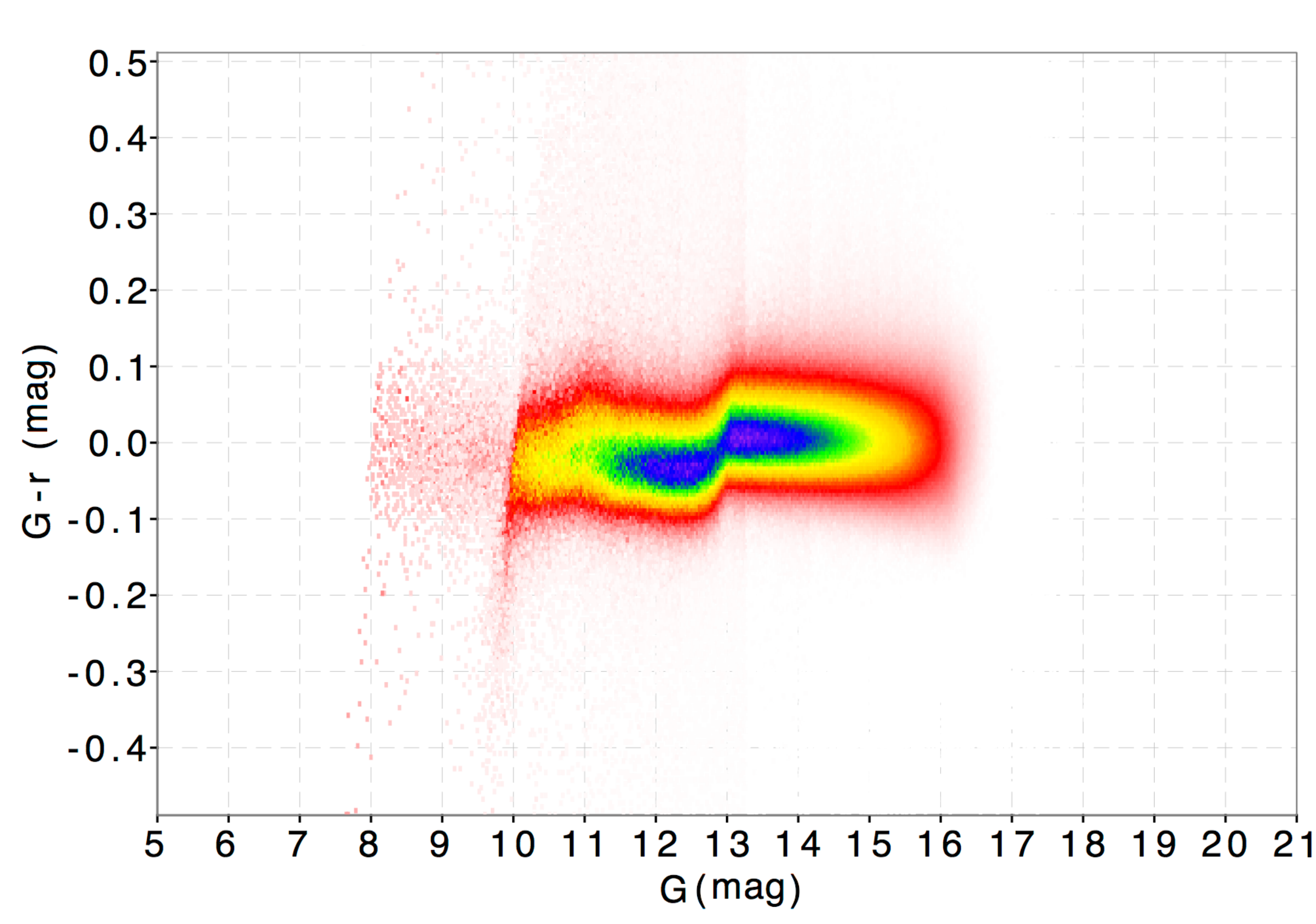}
   \includegraphics[width=0.4\textwidth]{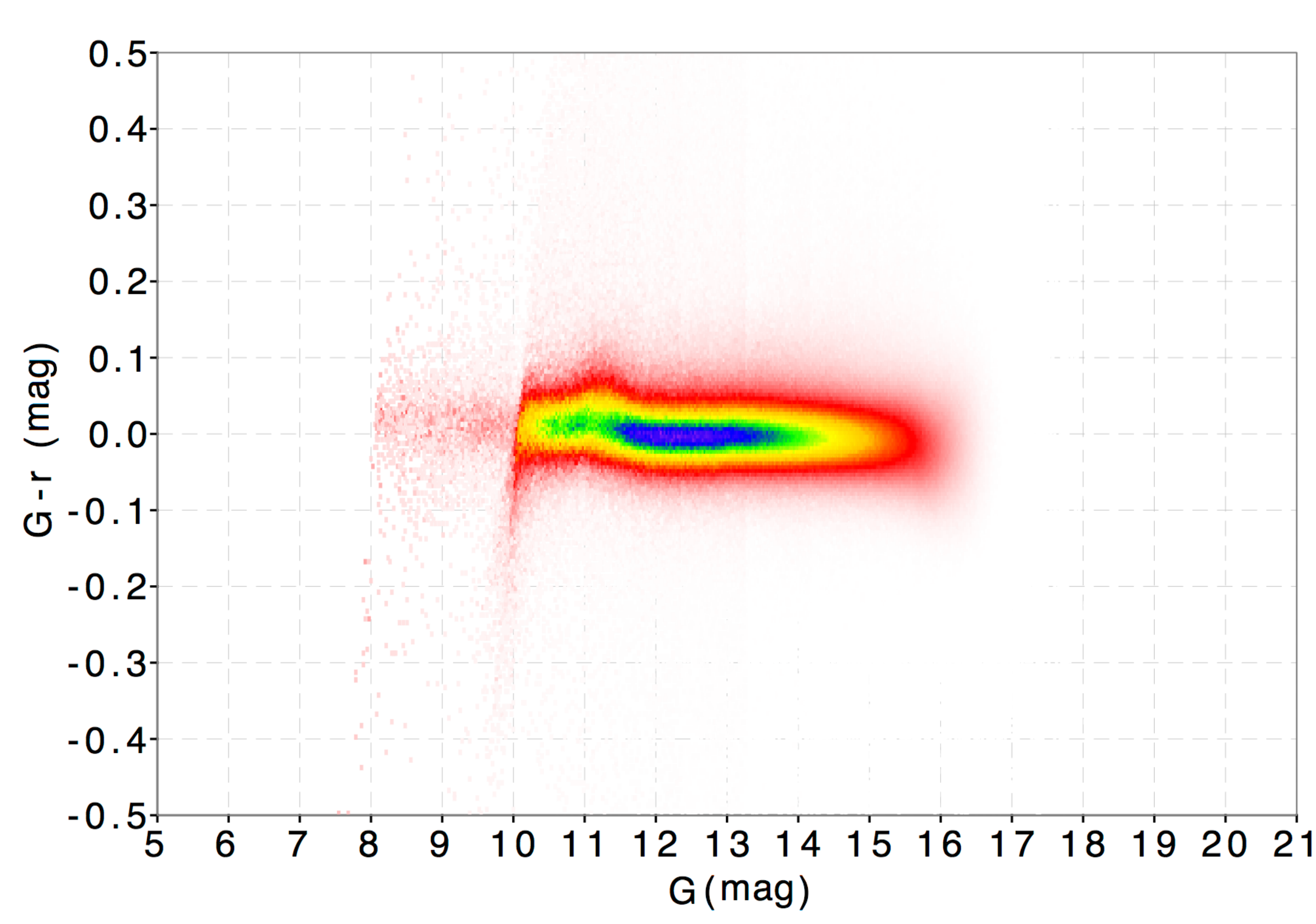}
   \end{center}  
	\caption{\small{Comparison between mean data derived from {\Gaia} observations and photometry from the APASS catalogue \citep{APASS} for raw
	(uncalibrated) transits before (top) and after (bottom) the gate/WC link calibration for a narrow range of colour.  The Galactic plane is excluded to minimize
extinction effects. The jump at $G=13$ in the top panel (and corrected in the bottom panel) is expected and due to the aperture correction effect not taken into account for
faint sources observed in 1D (see Sect.~\ref{sec:instrument}). The slight bump at $G=11$ is due to saturation and gating effects.
 }
\label{fig:kink}    
}
\end{figure}

The algorithm used for calibrating the systematic differences between 
CUs is a simple differential offset (zero point) calibration carried out 
using observations at either side of a boundary in magnitude for adjacent gates/WCs. 
A single link calibration was carried out for the entire {\GDR1} data time range since it was observed that the coefficients were constant over
this period. Since these calibrations are targeted at a specific gate/WC feature, they are effectively averaged over all
FoVs, rows and strips. Figure~\ref{fig:kink} compares the $G$ photometry with the APASS catalogue for a narrow range of
colours for different magnitudes before and after applying the link calibration. 
The offset at $G=13$ present in the uncalibrated data (top panel) is due to the poor mixing among 
different WCs. This is corrected when the link calibration is applied \citep[see also][]{PhotValidation}.

In order to apply these link calibrations, one gate was chosen as the reference and the differential offsets with respect to this reference gate were combined to form a
single functional form that covers the whole magnitude range. Since the link offsets are derived differentially, the process does not need reference fluxes. However, they
do need sources observed with more than one gate/WC combination. The derived offsets were applied to the raw fluxes before they were averaged for the first time to form the
initial set of reference fluxes. More details about how the link calibrations were used in the processing of the {\GDR1} data can be found in \cite{PhotProcessing}.

\section{Colour information for the internal calibrations} 
\label{sec:colour}

\newcommand{\fda}[1]{{\bf #1}}

\begin{figure} \resizebox{\hsize}{!}{\includegraphics[width=\hsize*4/9]{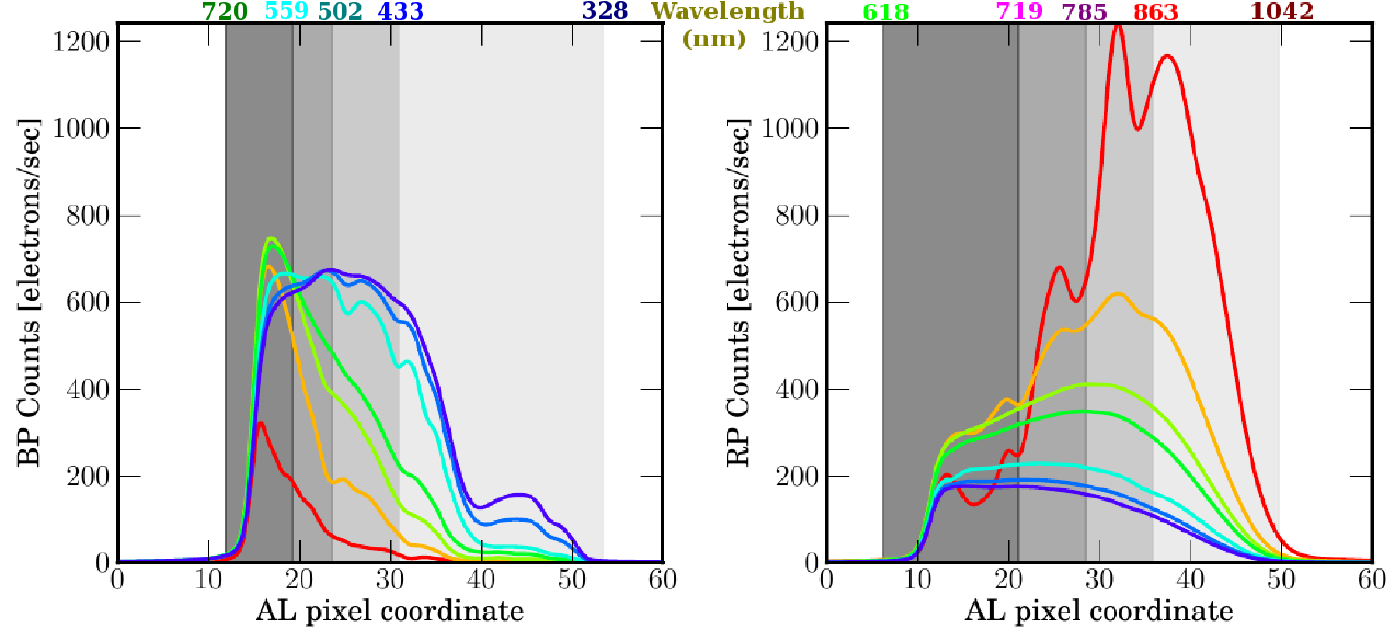}}
\caption{Definition of SSC bands (grey shaded areas), superimposed on simulated BP (left) and RP (right) spectra using BaSeL-2.2 \citep{basel} for a sample of sources with
effective temperatures ranging from $3000$ (red line) to $50\ 000$~K (blue line). Coloured numbers in the upper part of the plots show the wavelength limits detailed
in Table~\ref{tab:sscwl}.
\label{fig:ssc:definition}
}
\end{figure}

\begin{figure} \resizebox{\hsize}{!}{\includegraphics[width=\hsize*4/9]{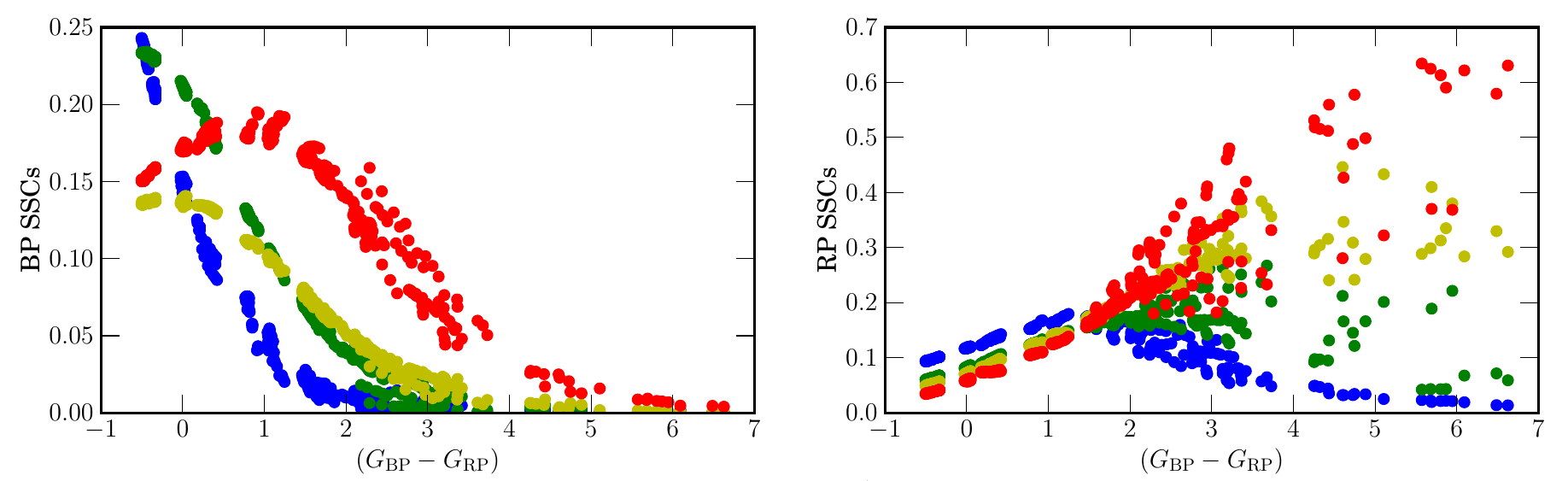}}
\caption{SSC dependency with $G_{\rm BP}-G_{\rm RP}$ colour for a range of noiseless simulated data using BaSeL-2.2  SEDs \citep{basel}.
\label{fig:ssc:expectation}
}
\end{figure}

The colour of the observed sources is required in many steps in the photometric
calibration and the PSF/LSF modelling.
This colour information required by the $G$ flux calibration
is obtained from the low-resolution spectra from the BP and RP spectrophotometers. As mentioned in Sect.~\ref{sec:internal}, SSCs are
used for the {\GDR1} calibrations. 
The wavelength ranges of the
SSC rectangular bands are given in Table \ref{tab:sscwl}. Figure~\ref{fig:ssc:definition} shows the location of the SSC bands in the
data space compared to the BP and RP simulated spectra for a sample of template stellar spectra covering the effective temperature
range from $3000$ to $50\ 000$~K.  Figure \ref{fig:ssc:expectation} shows the expected SSC dependency with the colour of the star for a
dataset of  simulated BP and RP spectra covering a wide range of spectral types.

\begin{table}
\begin{center}
\caption{Absolute wavelength boundaries for each rectangular SSC band.}
\tiny
\begin{tabular}{cc|cc}
\hline
\multicolumn{2}{c|}{BP} & \multicolumn{2}{c}{RP} \\
SSC id & $\lambda$ range (nm) & SSC id & $\lambda$ range (nm)\\
\hline
0 & [328, 433]& 4 & [618, 719]\\
1 & [433, 502]& 5 & [719, 785]\\
2 & [502, 559]& 6 & [785, 863]\\
3 & [559, 720]& 7 & [863, 1042]\\
\hline
\end{tabular}
\label{tab:sscwl}
\end{center}
\end{table}

Contrary to SM and AF observations, the BP/RP spectra enter the PhotPipe processing in their raw status.
Conversion to physical units and bias corrections are dealt with in the pre-processing stages. For a description of the bias 
correction we refer to the analogous treatment done for the SM and AF data described in \cite{IdtRef}, the main difference being 
that for BP/RP the full treatment, including the offset non-uniformities, is applied. 

SSC data are treated as fluxes and are calibrated in the same manner as 
the $G$, {\BP} and {\RP} fluxes. 
Future developments will include deriving the SSC values directly from the 
calibrated spectra, thus eliminating the need 
to 
 calibrate them separately.

The pre-processing of BP and RP spectra first requires a correction for the background flux contribution. The background model used for
BP and RP in {\GDR1} only takes into account the biggest contribution, which is the straylight, (see Sect.~\ref{sec:instrument}).  The model is
based on 2D maps in AC coordinates and spin angle with respect to the Sun (heliotropic angle).  For a more detailed description, see
\citealt{PhotProcessing}. When straylight is being subtracted, the smoother component of the astrophysical background is also removed.  The
charge release signal, caused by charge injections  introduced to mitigate the effects of CTI \citep{GaiaMission}, also contributes to the
background, but for {\GDR1}, this effect has not been calibrated and observations close to a charge injection have been filtered
out for this reason.

The pre-processing also involves the calibration of the geometry of the instrument. This is explained in more detail in the following
section.


The computation of raw SSCs from the observed spectra requires knowledge of the absolute dispersion function
providing the correspondence between data space coordinates and wavelength. Nominal BP and RP dispersion functions
have been used for {\GDR1}. These provide the location in sample space of a given wavelength with respect to the 
location of a reference wavelength. This reference position in the observed spectrum will in principle be different
for different transits for various reasons: i) sources may have a non-negligible AL 
motion, 
ii) the on-board window propagation based on the nominal geometry (derived from pre-launch measurements) may not 
be very accurate, and iii) the observation window is positioned using a grid with resolution equal to a macro-sample 
(group of four samples) which may result in additional misalignment.

The location of the reference wavelength in the window containing the dispersed image of the source can be predicted by extrapolation from the centroids of each of the AF
observations that precede the BP/RP observation in a FoV transit. This requires an accurate knowledge of the AL geometric calibration of the BP/RP CCDs with respect to the
AF values and of the satellite attitude. The BP/RP geometry is calibrated by comparing the predicted location of the source in the window with the actual location of the
reference wavelength in a set of observed spectra, selected to be of similar spectral type. For such a dataset, the data space coordinates of the reference wavelength are
expected to be the same, except for the effects of non-perfect centring of the window mentioned above.  Cross-correlation is first used to align the spectra after scaling
the fluxes to the same integrated flux. The adjustment in sample position and flux are then refined by fitting a spline to all spectra observed in the same CU and then
using that as a reference spectrum to fit second-order corrections to the alignment parameters.  After alignment, all spectra from different CUs are used to generate a
reference spectrum, which is then fitted back to each spectrum to evaluate the sample position of the reference wavelength within the actual sampling. The calibration
procedure is described in greater detail in \cite{PhotProcessing}. The difference between the predicted and actual locations is expressed in units of time. The observation
time (on-board mission timeline, OBMT; \citealt{GaiaMission}) is defined as the instant when the reference wavelength crosses the fiducial observation line defined for the
corresponding CCD, gate and FoV configuration.

The difference between the predicted observation time and the actual time is modelled \citep[see][]{LindegrenGDR1} as a function
$\eta(\mu)$ of the AC pixel  coordinate $\mu$, a continuous variable covering the 1966 pixel columns in a CCD. The value of $\eta(\mu)$ additionally depends
on a number  of parameters including the CCD index ($n$), FoV/telescope index ($t$), CCD gate ($g$), stitch block ($b$) and
time. The time dependency  is described by means of discrete calibration intervals. The length and boundary location for these intervals depend
on the  component being calibrated 
and on satellite events expected to affect the calibration parameters. 
The fit of the geometric calibration model depends on the AF geometric calibration via the predicted observation time which
is derived from the AF centroid as determined in IDT, the AF geometric calibration and the satellite attitude. 

The geometric calibration model is defined by different components:

\begin{equation}
\eta(\mu) = \eta^{\rm LS}_{nt}(\mu) + \eta^{\rm LS}_{g}(\mu) + \eta^{\rm SS}_{nb}(\mu)
\end{equation}

\begin{itemize} 
\item An LS component computed over a short timescale (approximately 5 days) defined by a linear 
combination of shifted Legendre polynomials describing overall effects of translation, rotation and curvature,
\begin{equation}
\eta^{\rm LS}_{nt}(\mu) = \sum_{l=0}^L \Delta\eta^{\rm LS}_{lnt} L^*_l (\tilde{\mu})
\end{equation}
where $L^*_l$ are shifted Legendre polynomials, orthogonal oat $[0, 1]$ and reaching $\pm1$ at the end points: $L^*_0(x)=1$, $L^*_1(x)=2x-1$, and
$L^*_2(x)=6x^2-6x+1$. Different sets of parameters are computed for each CCD ($n$)
and FoV ($t$) combination:
\item A simple offset for different gate configurations
$\eta^{\rm LS}_{g}(\mu) = \Delta\eta^{\rm
LS}_{g}$, computed  on a longer timescale, taking care of residual effects;
\item An offset for different CCD AC stitch blocks $\eta^{\rm SS}_{nb}(\mu) = \Delta\eta^{\rm
SS}_{nb}$ (see Sect.~\ref{sec:instrument}) also computed  on a long timescale, to account for effects due to the non-perfect manufacturing of the CCDs. The block index $b$
is defined as
$(\mu_b + 128.5)/250$, where $\mu_b$ are the block boundaries defined in Sect.~\ref{sec:instrument}.
\end{itemize}

The computed calibrations can be applied to all BP/RP spectra to estimate the location of the reference wavelength
in the data space. This allows the application of the nominal dispersion functions and thus convert the data space coordinate
into absolute wavelengths. 

It should be noted that for \GDR1 no calibration of the wavelength scale has been attempted. Nominal pre-launch
dispersion functions, based on an assumed reference wavelength and corresponding data space coordinate, were
used. The reference spectrum, created during the alignment procedure, could in principle have
an arbitrary offset with respect to the nominal value. This could result in an offset in the location of the SSC bands with
respect to the values given in Table \ref{tab:sscwl}.
The offset is expected to be small owing to the large number of spectra contributing to the reference spectrum definition, 
and consistent over the entire dataset.
A constant offset in the wavelength calibration would have no effect at all on the flux calibrations; it would simply imply a slightly different definition
of the SSC bands, but the SSCs would still be providing the colour information that is required for the calibration of the fluxes. 
A significant variation in time of the dispersion (which is  not expected) would also have no impact as it would  effectively produce a 
change in the passband, thus adding a time-dependent colour term which is already included in the calibration model definition.

\section{Reference photometry update} 
\label{sec:update}

At the start of each iteration (see Sect.~\ref{sec:principles} and Fig.~\ref{Fig:IterationPlot}), weighted mean fluxes are generated for each source. Weighted means were
chosen because the errors on each observation are not the same and depend on which observational configuration are used (see Sect.~\ref{sec:instrument}). Gating
configuration has the most significant effect since it determines the effective exposure time of the observation.

An inverse variance weighting scheme is used (i.e. $w_k=1/\sigma_k^2$, where $\sigma_k$ is the error on the flux quoted by the image parameter determination, IPD, see \citealt{IdtRef}). 

The weighted mean flux, $\overline{I_s}$ expressed in photoelectrons per second, for the source $s$ is then given by

\begin{equation}
\label{eq:reflux}
\overline{I_s}={\sum_{k} I_{sk} \frac{w_k}{\sum_{k} w_k}}
\end{equation}

\noindent where $I_{sk}$ are the individual photon flux measurements for observation $k$.

In the reductions used for {\GDR1}, the quoted errors for the single flux measurements, $\sigma_k$,
are not representative of the observed distribution of the fluxes since 
they do not account for model-fitting issues, such as using a non-representative PSF/LSF in the IPD fit. In {\GDR1} a simplified form of the PSF/LSF was used for the IPD that does not
include dependencies on AC motion and colour which are known to be needed\footnote{As part of the cyclic nature of the processing, these dependencies will be added to the reductions in
future data releases.}.
Thus, in order to estimate the error on the mean, it is necessary to account for any intrinsic scatter that might exist within the data caused by these effects.
The calculation used in this reduction is the same as that used for the Hipparcos catalogue \citep{Hipparcos}:

\begin{equation}
 \label{eq:sigmaIs}
   \sigma_{\overline{I_s}}=
       \sqrt{\sum_{k} I_{sk}^2 w_k - \overline{I_s}^2\sum_{k} \frac{w_k}{{N_{\rm obs}-1}}} \cdot
       {\frac{1}{\sqrt{\sum_{k} w_k}}}
\end{equation}

\noindent where $N_{\rm obs}$ is the number of single CCD observations. It is the weighted mean flux in Eq.~\ref{eq:reflux} and error in Eq.~\ref{eq:sigmaIs} that are used in the photometric calibrations as the reference flux.

This process is also carried out on the colour information (see Sect.~\ref{sec:colour}) and 
appropriate weighted mean fluxes are used for all sources in the application
of the calibrations to produce the calibrated photometr,y i.e. mean values rather than instantaneous values are used. 
This includes the integrated {\BP} and {\RP} fluxes and the SSC information.

At this stage variable sources cannot  yet be identified and they are handled in {\GDR1} in exactly the same way as constant sources, although their
colour can also be variable and epoch colours should be used in those cases. Future releases will be able to account
for their colour variability.

\def\fnu{f_{\nu,s}(\nu)}
\def\flamb{f_{\lambda,s}(\lambda)}
\def\ABZP{-56.090}

%
%
%

\section{External calibration}\label{sec:external}

The external calibration aims to derive the characteristics of the {\em mean}  
photometric instrument, where all possible differences and variations across the 
focal plane and time, etc, have been taken into account by the internal calibration process.

 \subsection{\texorpdfstring{$G$}{G} magnitude scale}
 
The weighted mean flux $\overline{I_s}$ provided by Eq.~\ref{eq:reflux} can be used to define an instrumental magnitude:   
   \begin{equation}
   \label{eq:Ginstr}
G_{\rm instr} = -2.5 \log  \overline{I_s} 
\end{equation}
  We define the {\Gaia} magnitude $G$ scale by adding a zero point, $G_0$, to the instrumental magnitude, as
\begin{equation}
\label{eq:gaiascale}
G \equiv G_{\rm instr} + G_0
\end{equation}

The aims of the external calibration are therefore to derive the shape of the mean system passband and the zero point $G_0$ to link the internally calibrated photometry to
an  
external (absolute) scale, specifically the VEGAMAG scale.

The shape of the passband can be obtained  by using an adequate parametric model and a 
suitable grid of calibrators (SPSS) whose absolute 
SEDs are known with great accuracy from ground-based observations (see the following section for details).
Once the passband is known, the zero point can be derived by comparison of the synthetic 
photometry $G_{\rm synth}$ and the corresponding {\Gaia} data $G_{\rm instr}$ for the SPSS. 
 The technique of calibration with synthetic photometry is well-known and has been described in many published papers, see e.g. \citet{bessell} for an exhaustive
discussion.
 
The reference spectrum chosen for the {\Gaia} photometric system is the same used for the SPSS calibration, i.e.
the CALSPEC spectrum \texttt{alpha\_lyr\_mod\_002} available from the public CALSPEC server\footnote{\url{http://www.stsci.edu/ftp/cdbs/current_calspec}}.
This flux table is a Kurucz model \citep{kurucz} for the Vega spectrum with  $T_{\rm eff} = 9400$~K,  $\log g =3.95$, and
$[M/H]=-0.5$
at R=500. 
The model has been normalised to the most recent STIS Vega flux distribution \citep{bohlin2004} at $554.5$--$557$~nm. 
We assume the visual magnitude for Vega is $V_{\rm Vega} = 0.023 \pm 0.008$~mag \citep{bohlin3}.

For the {\GDR}1 it was decided to skip the determination of the passband and 
assume instead the nominal pre-launch instrument response, which is modelled as the product 
of the following quantities derived by 
the industry 
during on-ground laboratory test campaigns: 
i) the telescope (mirrors) reflectivity; 
ii) the attenuation due to rugosity and molecular contamination of the mirrors; 
iii) the QE of the CCDs; and
iv) the prism (fused silica) transmittance curve (including filter coating on their surface) 
for the {\BP}/{\RP} case.
A plot of the nominal passbands as function of wavelength can be found in \cite{jordi2010}.

\begin{figure}[t!]
\resizebox{\hsize}{!}{\includegraphics[width=\hsize*4/9]{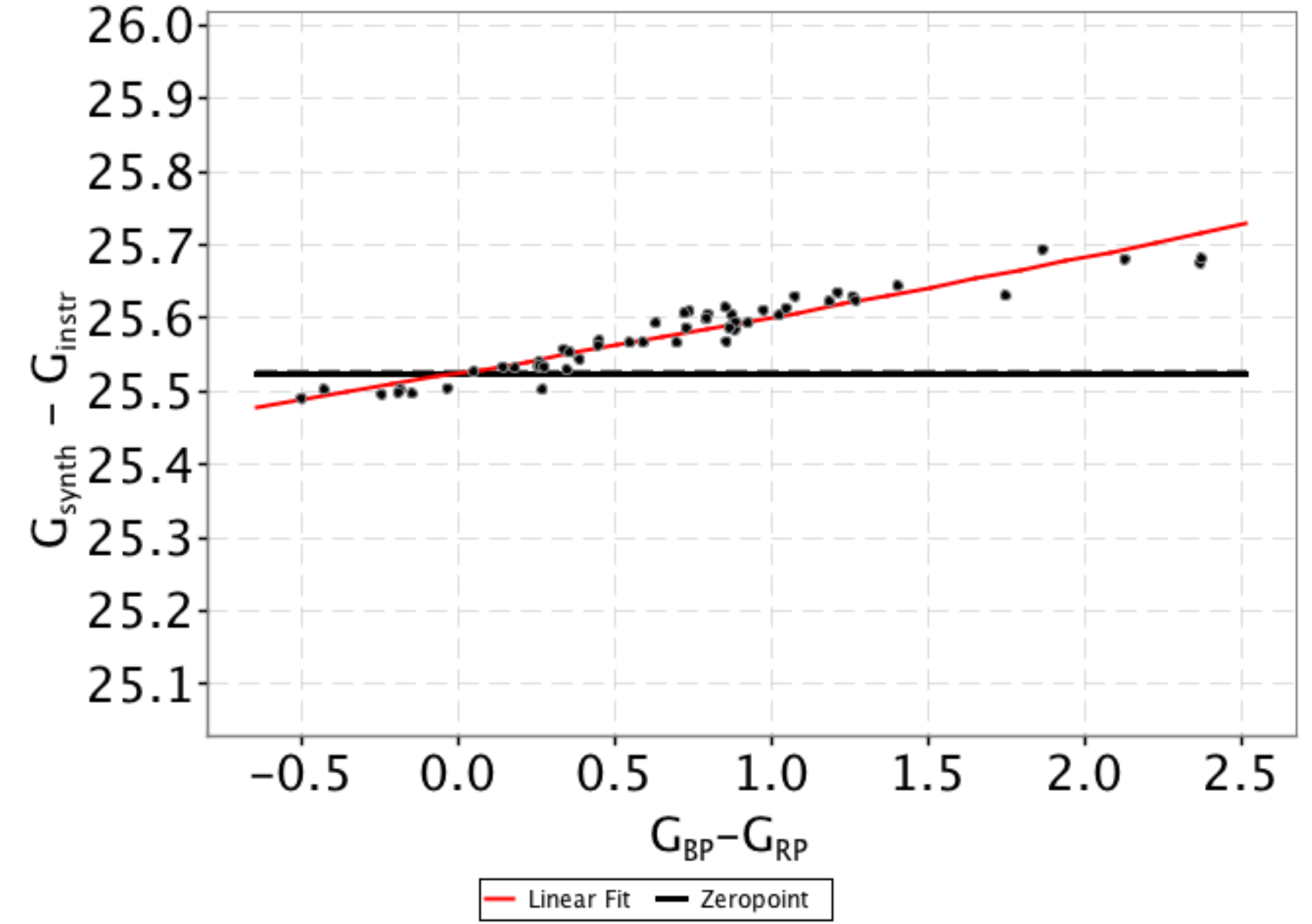}}
\caption{G band zero points $G_{\rm synth}-G_{\rm instr}$ vs 
{\BP}$-${\RP} colour, showing 
the colour term due to nominal passband usage in the \GDR1 calibration. 
The solid black line represents the zero point corresponding to zero colour. 
The red line shows the least-squares fit of the colour equation.
\label{fig:extcal}
}
\end{figure}

Since the nominal passband is expected to be somewhat different from the true one\footnote{The main difference is due to contamination, 
which mainly changes the response at the shortest wavelengths.}, the $(G_{\rm synth}-G_{\rm instr})$ vs. colour relation contains a  
colour term, as shown in  Fig. ~\ref{fig:extcal}. However, since the definition of the VEGAMAG system assumes that Vega 
has all colours 
equal to zero, the difference $(G_{\rm synth}-G_{\rm instr})$ must be independent from the passband for a 
source with zero colour. 
Therefore a calibration of the {\Gaia} data is still possible by deriving  the zero point 
for the G magnitude scale as the intercept of the $(G_{\rm synth} - G_{\rm instr})$ relation.

The value of the zero point in the VEGAMAG system is  $G_0 = 25.525 \pm 0.003$ mag (r.m.s. error)  
(more details on this calculation can be found in the online documentation\footlabel{rom}{\url{http://gaia.esac.esa.int/documentation/GDR1/index.html}}). A rough zero point
was also calculated for the AB system, and is $25.696 \pm 0.045$ mag (r.m.s. error), 
where the error is larger because the average value of $(G_{\rm synth} - G_{\rm instr})$  for all SPSS 
was used.

The external accuracy, 
estimated by comparison with a few data catalogues (Hipparcos, Tycho-2, Johnson), is presently 
of the order of 0.01-0.02 mag \citep{PhotValidation}, and is expected to improve in future data releases 
where the true passbands (also for {\BP} and {\RP}) will be used to derive the corresponding 
zero points.

\begin{figure}[!htbp]
  \begin{center}
   \includegraphics[width=0.49\textwidth]{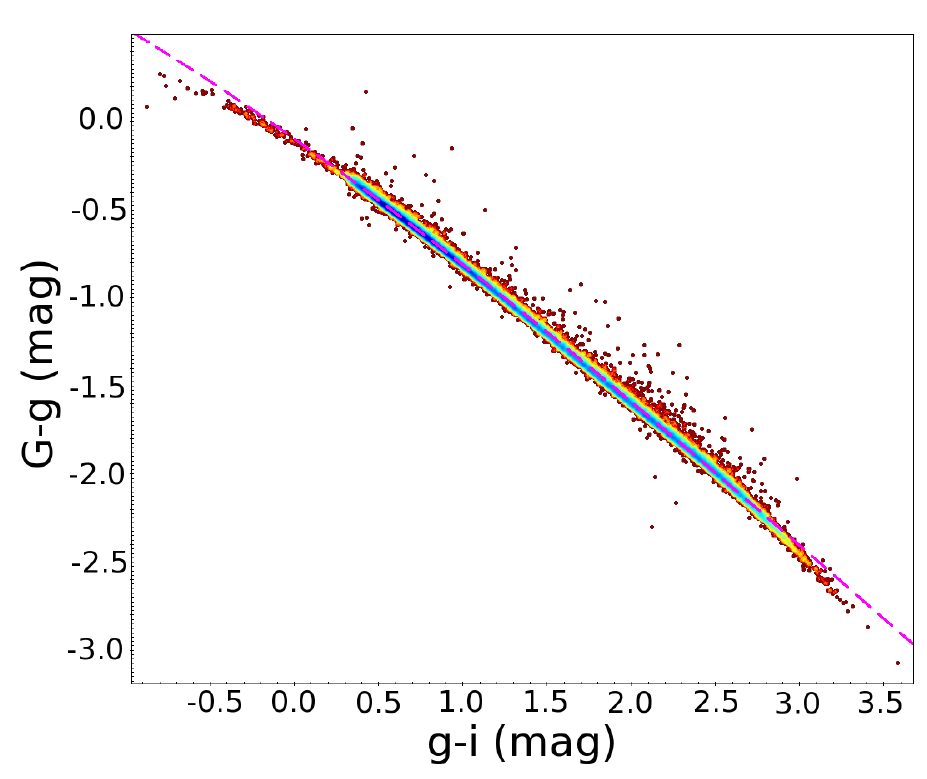} 
   \includegraphics[width=0.49\textwidth]{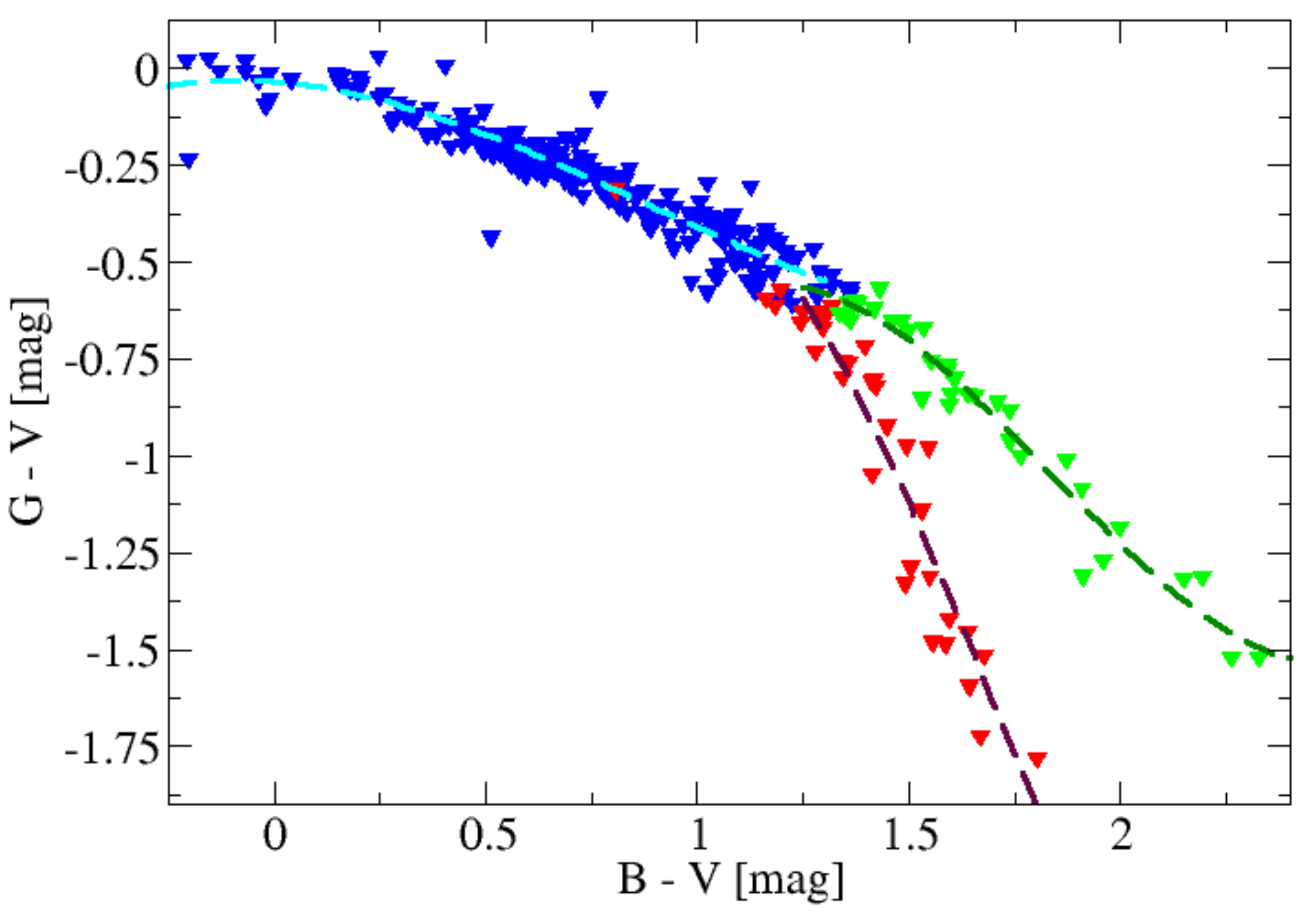} 
   \end{center}   
       \caption{\small{Fitting obtained between {\Gaia} and stars in SDSS stripe 82 (left) and Landolt standards (right) using {\GDR1} data.} 
\label{fig:PhotTransf}    
}   
\end{figure}

Photometric relationships were derived relating the {\Gaia} passband with other common photometric systems (see \citealt{PhotTopLevel} and {\GDR1} online
documentation\footref{rom}). The dispersion found in these derived relationships ranges between 0.03 and 0.12 mag, depending on the colours to be fitted\footnote{The
dispersion values specified here consider only those sources used to fit the photometric relationship after filtering out those sources with high uncertainties or that
deviate from the main behaviour.  Larger deviations due to cosmic dispersion can be found for some particular cases or when applied to different colour intervals. The
ranges of validity included in the {\GDR1} documentation must be carefully considered.}. Figure~\ref{fig:PhotTransf} shows two examples of these photometric relationships.
The left panel shows fitted polynomials between SDSS $g$ and $i$ passbands \citep{fukugita96} and the $G$ {\Gaia} passband. The right panel uses a subset of Landolt stars
\citep{landolt92} to derive a relationship between {\Gaia} and Johnson-Cousins $B$ and $V$ passbands. In this case, we have split the relationship into different colour
ranges: by using Tycho-Gaia Astrometric Solution (TGAS, \citealt{GDR1}) parallaxes, it was possible to separate red giants (green triangles) and dwarfs (red triangles)
and fit them separately to provide better results for the fitted law.

\subsection{Spectrophotometric standard stars (SPSS)}

The external calibration model requires the use of as many calibrators as possible (compatibly with the feasibility of the corresponding on-ground observing campaign),
including all spectral types from blue to red (to account for colour dependencies) with smooth SEDs, but also absorption features, both narrow (atomic lines) and wide
(molecular bands). The {\Gaia} end-of-mission requirement for the SPSS flux precision is $\simeq$1\%, and their flux calibration should be tied to Vega
\citep{bohlin2004,bohlin3,bohlin4} to within $\simeq$3\%. This establishes a requirement for approximately 200 calibrators, to ensure a homogeneous sky coverage all year
round in the northern and the southern hemispheres, and with a suitable magnitude range ($V\sim 9$--$15$~mag) to be observed by both {\Gaia} and several 2--4~m class
ground-based telescopes with a good S/N.

Because no existing set of SPSS in the literature simultaneously meets all these requirements, while at the same time covering the whole {\Gaia} spectral range
(330--1050~nm), an initial selection of approximately 300 SPSS candidates was made. These candidates cover all spectral types from hot white dwarfs, WD, and O/B to cold M
stars, i.e. with a temperature range {\teff}~$\sim 3500$--$80\ 000$~K.  A huge observational effort \citep{pancino2012,altavilla2015} to collect the required data and
monitor for constancy started in 2006 and was completed in 2015. The campaign was awarded more than 5000~hours of observing time, mostly in visitor mode, at six different
facilities: DOLORES at TNG in La Palma, EFOSC2 at NTT and ROSS@REM in La Silla, CAFOS at 2.2 m in Calar Alto, BFOSC at Cassini in Loiano, and LaRuca at 1.5 m in San Pedro
M{\'a}rtir. Some
additional data were also acquired with Meia at TJO in Catalonia.  The survey produced more than 100,000 imaging and spectroscopic frames that are presently being analysed
\citep{altavilla2015}. The raw data, flux tables, and intermediate data products are collected at the ASI Science Data Center\footnote{http://www.asdc.asi.it/} in a
database that will be opened to the public along with the first official release of SPSS flux tables\footnote{The first public release of SPSS flux tables should occur
before {\GDR2}.}.
\begin{figure}[t!]
\resizebox{\hsize}{!}{\includegraphics[width=\hsize*4/9]{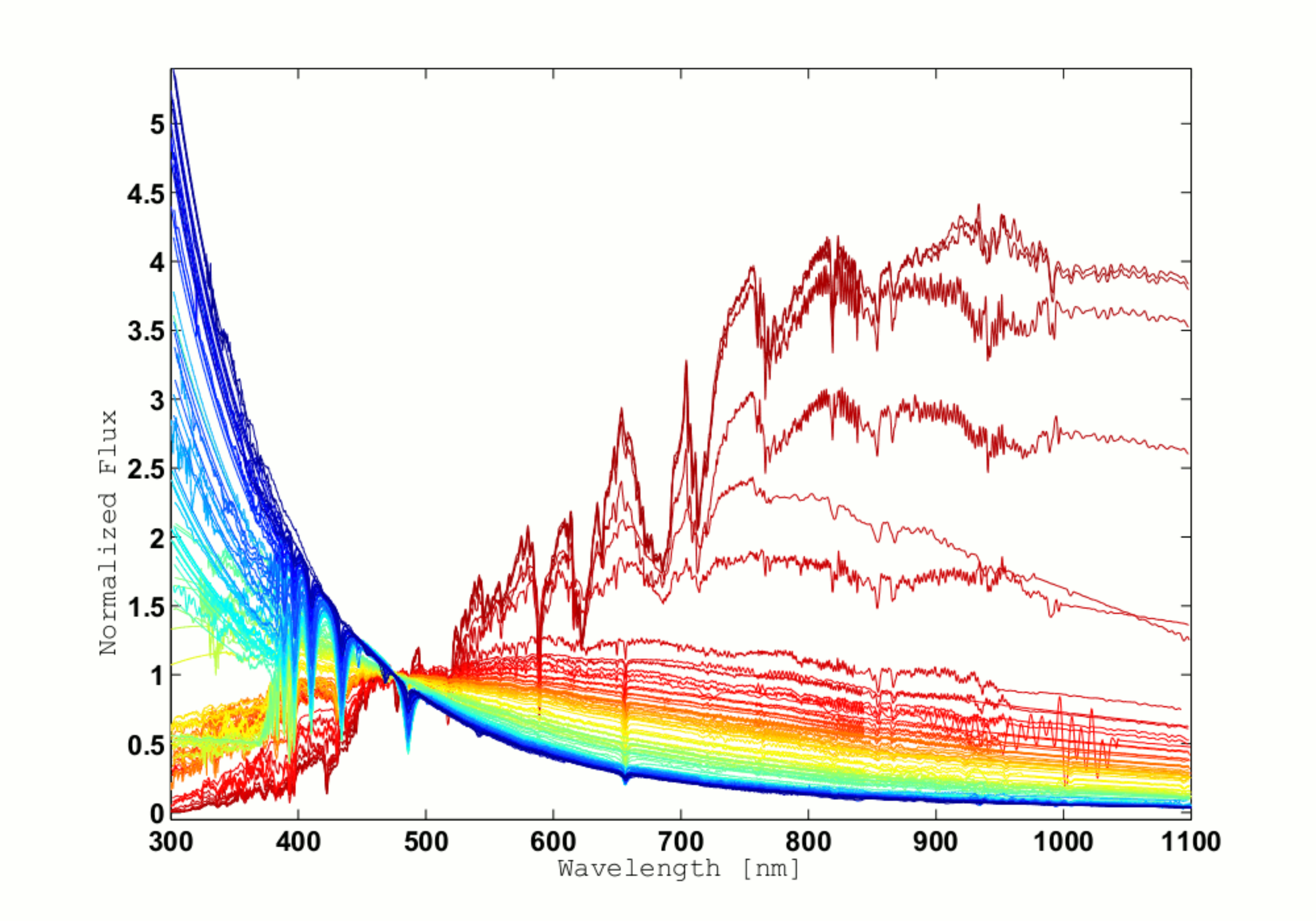}}
\caption{SPSS spectra ordered (and coloured) by spectral type, normalised in flux at 475 nm.
\label{fig:spss}
}
\end{figure}

Two internal releases of SPSS flux tables were prepared, a pre-launch version 
(V0) to test the instrument performance and the pipelines, and a first post-launch 
version (V1) to actually calibrate the photometry for {\GDR1}. Both V0 and V1 
contain the best 94 SPSS, i.e. about 50\% of the final SPSS sample, observed in
strictly photometric conditions and monitored for constancy on timescales of
1--2~hours to exclude stars with magnitude variations larger than $\pm$10~mmag
(Marinoni et al. 2016, in preparation). The quality of the flux tables in V1 already meets
the requirements (precision $<$1\% and accuracy $<$3\%). Future releases will 
increase the number of validated SPSS up to completion of the entire sample, 
improve the data quality of the flux tables, and complete the data products available 
for each SPSS (including magnitudes and variability assessment). Figure~\ref{fig:spss} shows a one-sight view 
of the current sample where fluxes are normalised at 475 nm, while the colour ranks with the sources spectral type.

\section{Summary}
\label{sec:conclusions}

This paper explains the basic principles of the photometric calibration model considered in {\GDR1}.

Owing to the 
variety of sources observed by {\Gaia} and to the complexity of the instrument, the photometric calibration is self-consistent, as it is not feasible 
to undertake on-ground campaigns for a set of thousands of standard stars evenly distributed in magnitude, colour and sky position.
For this reason, the calibration model is split into internal and external processes.

Internal calibration deals with the differential instrumental conditions to derive mean photometric values and to build a mean instrument using 
millions of constant
standard sources 
among all {\Gaia} observations. {\GDR1} uses all sources, even variable sources, as no variability analysis is
ready yet for {\GDR1}.

The internal calibration model for the $G$ passband accounts for overall response variations 
and depends on the colour of the source, the AC position in the focal plane, the telescope, CCD (row and strip) and observation time. 
In order to derive colour information, BP/RP spectrophotometric data, previously 
calibrated in geometry and background-subtracted, were used. 

In {\GDR1}, neither the aperture effect nor radiation damage was considered during the calibration process as they need input
information not yet fully available (or accurate enough) at this stage of the mission. These effects will be calibrated in future {\Gaia} data releases. 
Better calibrated PSF/LSF and the use of better colour information from BP/RP calibrated spectra will also contribute to 
improving the final
precision of the derived photometry. This first release was able to reach a typical accuracy of $3$~mmag (see
\citealt{PhotValidation}).

The internal calibration is done at two different scales. LS calibrations consider the global changes (mirrors and CCD QE), while SS calibration characterise in detail the
intra-CCD behaviour (groups of columns). Thanks to its larger spatial size, the LS calibration can be performed more often as it is able to collect observations faster than
for the SS CUs. To avoid the creation of different photometric systems for different magnitude ranges, a good mixing between different CUs is required. In cases when this
is not true, a link calibration is put in place.

Once an internal system is defined, common for all observed sources, the external calibration transforms the reference fluxes to
absolute physical units. External calibration relates the mean internal photometry with the absolute fluxes derived
using a 
set of standard stars (94 SPSS were used in {\GDR1} to be extended to about $200$ in future releases). In
{\GDR1}, this external calibration 
derives
a zero point and photometric transformations to allow users to
compare this data with existing photometry (see {\GDR1} online documentation\footref{rom}). 

\begin{acknowledgements}
We want to thank the anonymous referee who helped improve our first submitted version of the paper making it more complete and clearer to read.
This work was supported in part by the MINECO (Spanish Ministry of Economy) - FEDER through grant ESP2013-48318-C2-1-R and MDM-2014-0369 of ICCUB (Unidad de Excelencia 'Mar\'ia de Maeztu').
We also thank the Agenzia Spaziale Italiana (ASI) through grants ARS/96/77, ARS/98/92, ARS/99/81, I/R/32/00, I/R/117/01, COFIS-OF06-01, ASI I/016/07/0, ASI I/037/08/0, ASI I/058/10/0, ASI 2014-025-R.0, ASI 2014-025-R.1.2015,
and the Istituto Nazionale di AstroFisica (INAF).

This work has been supported by the UK Space Agency, the UK Science and Technology Facilities Council.

The research leading to these results has received funding from the 
European Community's Seventh Framework Programme (FP7-SPACE-2013-1) 
under grant agreement no. 606740.

The work was supported by the Netherlands Research School for Astronomy (NOVA) and the Netherlands Organisation for Scientific Research (NWO) through
grant NWO-M-614.061.414.
\end{acknowledgements}

%
%

\bibliographystyle{aa} 

\end{document}